\documentclass[prd,twocolumn,superscriptaddress,showpacs,nofootinbib,preprintnumbers]{revtex4}

\usepackage{amsmath}
\usepackage{graphicx,bm}
\usepackage{dcolumn}    

\begin{document}
\title{Exact Wave Propagation in a Spacetime with a Cosmic String}
\author{Teruaki Suyama}
\affiliation{Department of Physics, Kyoto University, Kyoto 606-8502, Japan}

\author{Takahiro Tanaka}
\affiliation{Department of Physics, Kyoto University, Kyoto 606-8502, Japan}

\author{Ryuichi Takahashi}
\affiliation{Division of Theoretical Astrophysics, National Astronomical Observatory of Japan, Mitaka, Tokyo 181-8588,Japan}

\preprint{KUNS-1997}
\date{today}

\begin{abstract}
We present exact solutions of the massless Klein-Gordon equation in 
a spacetime in which an infinite straight cosmic string resides.
The first solution represents a plane wave entering perpendicular to the
string direction. We also present and analyze a solution with a static 
point-like source. 
In the short wavelength limit these solutions 
approach the results obtained by using 
the geometrical optics approximation:
magnification occurs if the observer lies in front of the string 
within a strip of angular width $8\pi G\mu$,
where $\mu$ is the string tension.
We find that when the distance from the observer 
to the string is less than 
$ 10^{-3} {(G \mu)}^{-2}\lambda \sim 150 {\rm Mpc} (\lambda/{\rm AU}) (G\mu/10^{-8})^{-2}$,
where $\lambda$ is the wave length, 
the magnification is significantly reduced compared 
with the estimate based on the geometrical optics 
due to the diffraction effect. 
For gravitational waves from neutron star(NS)-NS mergers 
the several lensing events per year may be detected by DECIGO/BBO. 
\end{abstract}


\maketitle

\section{Introduction}
Typical wavelength of gravitational waves from astrophysical compact
objects such as BH(black hole)-BH binaries is in some cases 
very long so that 
wave optics must be used instead of geometrical optics when 
we discuss gravitational lensing. 
More precisely,
if the wavelength becomes comparable or longer than the
Schwarzschild radius of the lens object,
the diffraction effect becomes important and as a result 
the magnification factor approaches 
unity \cite{Ohanian,Bliokh,Bontz,Thorne,Deguchi}.
Mainly due to the possibility that the wave effects could be
observed by future gravitational wave observations,
several authors \cite{Takahashi:2003ix,Seto:2003iw,Nakamura:1997sw,Yamamoto:2003cd,Baraldo:1999ny,T.Nakamura:1999,Yamamoto:2003wg,Suyama:2005mx,Takahashi:2005sx,Takahashi:2004mc}
have studied wave effects in gravitational lensing in recent years.

In most of the works which studied gravitational lensing phenomenon
in the framework of wave optics, isolated and normal 
astronomical objects such as galaxies are concerned as lens objects.
Recently Yamamoto and Tsunoda\cite{Yamamoto:2003wg} studied wave 
effects in gravitational lensing by an infinite straight cosmic string. 
The metric around a cosmic string is completely different from
that around a usual massive object. 

Cosmic strings generically arise as solitons in a grand unified 
theory and could be produced in the early universe as a result 
of symmetry breaking phase transition\cite{Hindmarsh:1994re,vilenkin}. 
If symmetry breaking occurred after inflation, 
the strings might survive until the present universe.  
Recently, cosmic strings attract a renewed interest 
partly because a variant of their formation mechanism 
was proposed in the context of the brane inflation
scenario\cite{Dvali:1998pa,Dvali:1999tq,Burgess:2001fx,Alexander:2001ks,Dvali:2001fw,Jones:2002cv,Shiu:2001sy}. 
In this scenario inflation is driven by the attractive force
between parallel D-branes and parallel anti D-branes
in a higher dimensional spacetime. 
When those brane-anti-brane pairs collide and annihilate 
at the end of inflation, 
lower-dimensional D-branes, which behave like monopoles,
cosmic strings or domain walls from the view point of 
four-dimensional observers, are formed generically 
\cite{Majumdar:2002hy,Dvali:2002fi,Jones:2003da,Dvali:2003zj,Copeland:2003bj}.

For some time,
cosmic string was a candidate for the seed of structure
formation of our universe, but this possibility was ruled out 
by the measurements of the spectrum of  
cosmic microwave background (CMB)
anisotropies\cite{Spergel:2003cb,Percival:2002gq}.
The current upper bound on the dimensionless string tension 
$G\mu$ is around $10^{-7} \sim 10^{-6}$, which comes from the 
observations of CMB\cite{Jeong:2004ut,Pogosian:2003mz,Pogosian:2004ny,Wyman:2005tu}
and/or the pulsar timing \cite{Kaspi:1994hp,Thorsett:1996dr,McHugh:1996hd,Lommen:2002je}. 
Although cosmic string cannot occupy dominant fraction of the
energy density of the universe,
its non-negligible population is still allowed observationally\cite{Bouchet:2000hd,Rocher:2004my}. 
In fact, Sazhin et al.\cite{Sazhin:2005fd,Sazhin:2003cp} reported
that CSL-1, 
which is a double image of elliptical galaxies with angular separation 
$ 1.9~{\rm arcsec} $, could be the first case of 
the gravitational lensing 
by a cosmic string with $ G\mu \approx 4\times 10^{-7} $.

We study in detail wave effects in the gravitational lensing by an
infinite straight cosmic string.
In Ref.~\cite{Yamamoto:2003wg},
wave propagation around a cosmic string was studied but
they put the waveform around the string by hand. 
\footnote{After submitting this paper, we have noticed a paper \cite{Linet:1986db} in which the solutions of the wave equations around the cosmic string are given, though the apparent expressions are different from those given in this paper. In \cite{Linet:1986db} the author estimated the amplitude of the diffracted wave to be suppressed by ${\cal O}(G\mu)$ compared with that corresponding to the geometrical optics. We show that the importance of the diffraction effects are determined by the combination of three parameters, $G\mu$, the distance from the string to the observer and the wavelength and that the relative amplitude of the diffracted wave can be ${\cal O}(1)$ for realistic astrophysical situations.} 
Their prescription is correct only in the limit of geometrical optics,  
which breaks down when the wavelength becomes longer than a certain
characteristic length.
In this paper, 
we present exact solutions of the (scalar) wave equation 
in a spacetime with a cosmic string.
We analytically show that our solutions reduce to the results of the
geometrical optics in the short wavelength limit.
We derive a simple analytic formula of the leading order 
corrections to the geometrical optics 
due to the finite wavelength effects 
and also an expression for the long wavelength limit.
Interference caused by the lensing remains 
due to the diffraction effects even when 
only a single image can be seen in the geometrical optics. 
This fact increases the lensing probability by cosmic strings.

This paper is organized as follows.
In section II,
we construct a solution of the wave equation on a background spacetime
with an infinite straight cosmic string in the case that 
a source of the wave is located infinitely far. 
An extension to the case in which a point source is located at a finite 
distance is given in Appendix B. 
In section III,
we study 
properties of the solution obtained in sec. II in detail.
In section IV,
we focus on compact binaries as the sources of gravitational waves
and discuss the possible effects due to finiteness of 
the lifetime and the frequency evolution of the binaries
on the detection of the gravitational waves which pass near 
a cosmic string. 
We also give a rough estimate for the event rate of the lensing of 
gravitational waves from NS-NS mergers assuming DECIGO/BBO.
Section V is devoted to summary.

\section{A solution of the wave equation around an infinite straight cosmic string}
A solution of Einstein equations around an infinite straight
cosmic string to first order in $ G \mu $ is given 
by \cite{Vilenkin:1981zs}
\begin{equation}
d^2 s= -dt^2+dr^2+{(1-\Delta)}^2 
r^2 d {\theta}^2+dz^2, \label{1.1} 
\end{equation}
where $ (r, z, \theta) $ is a cylindrical coordinate($ 0 \le \theta < 2
\pi $) and $2\pi\Delta\approx 8\pi G\mu$ 
is the deficit angle around 
the cosmic string. 
Spatial part of the above metric 
describes the Euclidean space with a
wedge of angular size $2\pi\Delta $ removed. 
Due to the deficit angle around a string,
double images of the source are observed with an angular separation 
$\alt 2 \pi \Delta $ 
when a source is located behind the string in the limit of geometrical
optics. In general for a wave with a finite wavelength, 
some interference pattern appears.  
An exact solution of Einstein equations around a 
finite thickness string has 
been already obtained \cite{Gott:1984ef}, 
but we use the metric (\ref{1.1}) as a background since 
the string thickness is negligibly small compared with the 
Einstein radius, $\approx \pi D\Delta  $, where $ D $ is the distance 
from the observer to the string.

Throughout the paper, 
we consider waves of a massless scalar field instead of 
gravitational waves for simplicity, but 
the wave equations are essentially the same in these two cases.
An extension to the cosmological setup 
is straightforwardly done by adding an overall scale factor. 
In that case the time coordinate $t$ is to be understood as 
the conformal time. The wave equation remains unchanged
if we consider a conformally coupled field, but it is 
modified for the other cases due to curvature scattering. 
The correction due to  curvature scattering of the Friedmann 
universe is suppressed by the square of the ratio 
between the wavelength and the Hubble length, which can be
neglected in any situations of our interest. 

Our goal of this section is to construct a solution of the
wave equation which corresponds to a plane wave injected 
perpendicularly to and scattered by the cosmic string.
This situation occurs if the distance between the source and the 
string is infinitely large. 
In order to construct such a solution,
we introduce a monochromatic source uniformly extended in 
the $z$-direction and localized in $r-\theta$ plane, 
\begin{equation}
S=\frac{B}{(1-\Delta)} \delta (r-r_o) \delta (\theta-\pi) e^{-i\omega t}, \label{1.2}
\end{equation}
where $\omega$ is the frequency and 
we have introduced $B$, 
a constant independent of $\Delta$, 
to adjust the overall normalization 
when we later take the limit $r_o\to\infty$. 
The factor ${(1-\Delta)}^{-1}$ appears because $\theta$-coordinate 
used in the metric (\ref{1.1}) differs from the usual angle 
\begin{equation}
\varphi\equiv (1-\Delta)\theta. \label{1.25}
\end{equation} 
Here we consider a 
uniformly extended source instead 
of a point source since the former is easier to handle. 
When the limit $r_o\to\infty$ is taken, 
the answers are identical in these two cases. 
The case with a point-like source at a finite distance 
is more complicated.  
This case is treated in Appendix B.
 
Now the wave equation that we are to solve is
\begin{eqnarray}
 && \left( \frac{ {\partial}^2 }{\partial r^2}+\frac{1}{r}
 \frac{\partial}{\partial r}+\frac{1
}{(1-\Delta)^{2} r^2} \frac{
 {\partial}^2 }{\partial {\theta}^2} +{\omega}^2 \right) \phi
(r,\theta)\cr
 &&\qquad\qquad  =\frac{B}{1-\Delta} \delta (r-r_o) \delta (\theta-\pi).
 \label{1.3}
\end{eqnarray}
Since $\phi (r,-\theta)$ satisfies 
the same equation~(\ref{1.3}) as $\phi(r,\theta)$ does, 
$\phi(r,\theta)$ is even in $\theta$. 
Thus, it can be expanded as
\begin{equation}
\phi(r,\theta)=\sum_{m=0}^{\infty} f_m (r) \cos m \theta. \label{1.4}
\end{equation}
From Eqs.~(\ref{1.3}) and (\ref{1.4}),
the equations for $f_m(r)$ are
\begin{eqnarray}
&& \left( \frac{d^2}{dr^2}+\frac{1}{r} \frac{d}{dr}+{\omega}^2 -\frac{
 {\nu}^2_m }{r^2} \right) f_m (r)\cr
 &&\qquad ={\epsilon}_m \frac{{(-1)}^m}{1-\Delta}
\frac{B}{2\pi} \delta (r-r_o), \label{1.5}
\end{eqnarray}
where ${\epsilon}_o \equiv 1, {\epsilon}_m \equiv 2 (m \ge 1) $ and ${\nu}_m \equiv {(1-\Delta)}^{-1} m$.
The solution of Eq.~(\ref{1.5}) except for $r=r_o$ is a linear combination
of Bessel function and Hankel function.
We impose that the wave $\phi$ is regular at $r=0$ and pure out-going
at infinity.
Further, imposing that the wave is continuous at $r=r_o$,
$f_m(r)$ becomes
\begin{eqnarray}
f_m(r)=N_m \left( H^{(1)}_{{\nu}_m}(\omega r_o) J_{ {\nu}_m}(\omega r) \Theta (r_o-r) \right. \nonumber \\
\left. +J_{{\nu}_m}(\omega r_o) H^{(1)}_{{\nu}_m}(\omega r) \Theta (r-r_o) \right), \label{1.6}
\end{eqnarray}
where $\Theta(x)$ is the Heaviside step function.
Substituting Eq.~(\ref{1.6}) into Eq.~(\ref{1.5}),
the normalization factor $N_m$ is determined as 
\begin{eqnarray}
N_m&=&\frac{B}{1-\Delta} \frac{ {\epsilon}_m {(-1)}^m}{2\pi\omega} \nonumber \\
&&\times \left[
    J_{{\nu}_m}(\omega r_o) H^{(1)'}_{{\nu}_m}(\omega r_o)
  -H^{(1)}_{{\nu}_m} (\omega r_o) J'_{{\nu}_m}(\omega r_o)
  \right]^{-1}\cr
& =&{Br_0 \epsilon_m(-1)^m\over 4i(1-\Delta)},
 \label{1.7}
\end{eqnarray}
where $'$ denotes a differentiation with respect to the argument.
From Eqs.~(\ref{1.6}) and (\ref{1.7}) with the aid of the 
asymptotic formulae of the Bessel and Hankel functions,
$\phi(r,\theta)$ for $r_o \to \infty$ can be written as
\begin{eqnarray}
\phi(r,\theta)=\frac{-iB}{2\sqrt{2}(1-\Delta)} \sqrt{\frac{r_o}{\pi \omega}} 
  e^{i\omega r_o-i \frac{\pi}{4}} \nonumber \\
 \times \sum_{m=0}^{\infty} {\epsilon}_m i^m
e^{-\frac{i m \pi\Delta }{2(1-\Delta)}} 
  J_{{\nu}_m}(\omega r) \cos
 m\theta. \label{1.8}
\end{eqnarray} 
We determine the overall normalization of the source amplitude $B$, 
independently of $G\mu$, 
so that Eq.~(\ref{1.8}) becomes a plane wave 
$e^{i \omega r \cos \theta}$ when $G\mu=0$. 
This condition leads to 
$B=-2\sqrt{\frac{2\pi \omega}{r_o}}e^{-i\omega r_o-i\pi/4}$. 
Then, finally $\phi$ becomes
\begin{equation}
\phi(r,\theta)=\frac{1}{1-\Delta} \sum_{m=0}^{\infty} {\epsilon}_m i^m
 e^{- \frac{im \Delta\pi}{2(1-\Delta)}} J_{{\nu}_m}(\omega r) \cos
 m\theta. \label{1.9}
\end{equation}

\section{Limiting behaviors of the solution}
\label{sec:behaviorsofsolution}
\subsection{Approximate waveform in the wave zone}
The solution~(\ref{1.9}) describes the waveform
propagating around a cosmic string.
But it is not easy to understand the behavior of the solution 
because it is given by a series. In fact, it takes much
time to perform the summation in Eq.~(\ref{1.9}) numerically for
a realistic value of tension of the string, say, 
$G\mu \lesssim 10^{-6}$ because of 
slow convergence of the series. 
In particular it is not manifest whether 
the amplification of the solution 
in the short wavelength limit coincides with the one which is obtained
by the geometrical optics approximation.
Therefore it will be quite useful if one can derive a simpler 
analytic expression. 
Here we reduce the formula by assuming that the distance between 
the string and the observer is much larger than the wave length,
\begin{equation}
\xi \equiv \omega r\gg 1,
\end{equation}
which is valid in almost all interesting cases.

\begin{figure}[h]
\begin{center}
  \includegraphics[width=6.cm,clip]{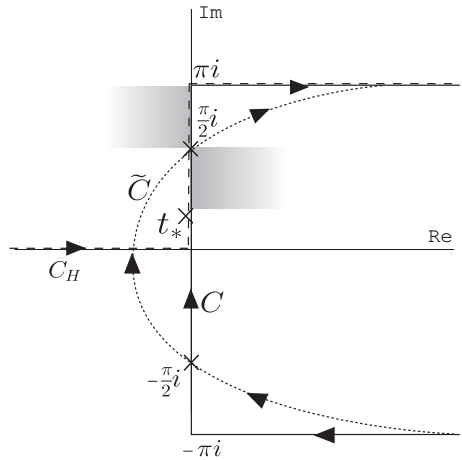}
\caption{Black, dotted and dashed lines are contours of the 
integral $C$, ${\tilde C}$ and $C_H$,
respectively.
$\pm i \frac{\pi}{2}$ are the saddle points of $e^{\xi \sinh t}$.
}
\label{contour}
\end{center} 
\end{figure}

Using an integral representation of the Bessel function,
\begin{equation}
J_{\nu}(\xi)=\frac{1}{2i\pi} \int_C dt \
e^{\xi \sinh t-\nu t}, \label{3.1}
\end{equation}
where the contour of the integral $C$ is such 
as shown in Fig.~\ref{contour},
Eq.~(\ref{1.9}) can be written as
\begin{eqnarray}
\phi (\xi, \theta)\!\!&=&\!\!-\frac{J_0 (\xi)}{1-\Delta}+\frac{1}{1-\Delta} 
 \frac{1}{2i\pi} \int_C dt~ e^{\xi \sinh t} \nonumber\\
&\times& \!\!\sum_{m=0}^{\infty} e^{-\frac{mt}{1-\Delta}
 +\frac{\pi}{2}mi-\frac{i m \pi\Delta}{2(1-\Delta)}}
 (e^{im\theta}+e^{-im\theta}). \label{3.2}
\end{eqnarray}
When $t$ is in the segment of the integration contour $C$ 
along the imaginary axis, 
the summation over $m$ does not converge 
because the absolute value of each
term in the summation is all unity.
In order to make the series to converge, 
we need to think that the integration contour $C$ is not 
exactly on the imaginary axis but $t$ always has 
a positive real part. 
For bookkeeping purpose, 
we multiply each term in the sum by a factor $e^{-\epsilon m}$
($\epsilon$ is an infinitesimally small positive real number).
Then Eq.~(\ref{3.2}) becomes
\begin{eqnarray}
\phi (\xi, \theta)=-\frac{J_0 (\xi)}{1-\Delta}+\psi (\xi, \theta)+\psi
 (\xi, -\theta), \label{3.31}
\end{eqnarray}
where $\psi (\xi,\theta)$ is defined by 
\begin{eqnarray}
\psi (\xi,\theta):=\frac{1}{1-\Delta} \frac{1}{2i\pi} \int_C dt~
\frac{e^{\xi \sinh
t}}{1-e^{-\frac{t-t_\ast}{1-\Delta}}}, 
\label{3.3}
\end{eqnarray}
with
\begin{equation}
t_{\ast}:=-\epsilon+i\frac{\pi}{2}-i\frac{{\alpha}(\theta)}{\sqrt{\xi}}, 
\label{3.3a}
\end{equation}
and ${\alpha}(\theta):= (\pi \Delta -(1-\Delta) \theta) \sqrt{\xi}$.

Now we find that
all we need to evaluate 
is $\psi (\xi,\theta)$ in order
to obtain an approximate formula for $\phi (\xi,\theta)$. 
This integral will not be expressed by simple known functions
in general, but the integration can be performed by using 
the method of steepest descent in the limit $\xi\gg 1$. 

The integrand of Eq.~(\ref{3.3}) has two saddle points  
located at $t=t_+\approx i\pi/2$ and 
$t=t_-\approx -i\pi/2$ in the vicinity of 
the integration contour $C$. 
We should also notice that 
the integrand has a pole at $t=t_{\ast}$, which is also 
infinitesimally close to the contour of the integral $C$. 
This pole is located near the saddle point at $t=t_+$ 
as far as $\Delta$ and $\theta$ are small.  
Hence the treatment of the saddle point at $t=t_+$ is 
much more delicate than that of the saddle point at $t=t_-$.  
We only discuss the saddle point at $t=t_+$, then the case 
at $t=t_-$ is a trivial extension. 
 
When $\Re (t)>0, \Im (t)<i\frac{\pi}{2}$ or 
$\Re (t)<0, \Im (t)>i\frac{\pi}{2}$, 
which corresponds to shaded regions in Fig.~\ref{contour},
$e^{\xi \sinh t}$ diverges in the limit $\xi \to \infty$.
If ${\alpha}(\theta)>0$, 
the pole at $t=t_{\ast}$ is in the bottom-left unshaded region. 
In this case we cannot deform the contour to the 
direction of the steepest descent at $t=t_+$ 
without crossing the pole at $t=t_{\ast}$. 
The deformed contour which is convenient to apply the method 
of the steepest descent is such that is shown 
as ${\tilde C}$ in Fig.~\ref{contour}. 
When we deform the integration contour from $C$ to $\tilde C$,
there arises an additional contribution corresponding to the residue at 
$t=t_\ast$ when ${\alpha}(\theta)>0$. 
On the other hand, 
if ${\alpha}(\theta) <0$, 
the pole is in the top-left shaded region. 
In this case, we can deform the contour of the integral to 
the direction of the steepest descent 
without crossing the pole $t_{\ast}$. 
Hence no additional term arises. 

From these observations,
we find that it is necessary to evaluate the integral~(\ref{3.3}) 
separately depending on the signature of ${\alpha}(\theta)$. 
Though the calculation itself can be done straightforwardly,  
it is somewhat complicated because 
the saddle point and the pole are close to each other. 
When the pole is located inside the region around the saddle 
point that contributes dominantly to the integral, a 
simple Gaussian integral does not give a good approximation. 
Detailed discussions about this point are given in Appendix A. 
Here we only quote the final result which keeps terms up to 
$O(1/\sqrt{\xi})$, 
\begin{eqnarray}
\psi(\xi,\theta) &\approx &\exp \left( i\xi \cos \frac{
{\alpha}(\theta)}{\sqrt{\xi}} \right) \Theta
({\alpha}(\theta))\cr
&& -\frac{{\sigma}(\theta)}{\sqrt{\pi}} 
  \exp \left( i\xi+\frac{i {\alpha}(\theta)}{
   (1-\Delta) \sqrt{\xi}}-\frac{i}{2}
   {\tilde\alpha}^2(\theta) \right)\cr 
&&\quad\times {\rm Erfc} 
  \left( {\sigma}(\theta) \frac{{\tilde\alpha}(\theta) }
    {\sqrt{2} } e^{-i
   \pi/4} \right) \cr
&& + \frac{1}{\sqrt{2\pi \xi}} \frac{1}{1-\Delta} 
     \frac{ e^{-i \xi +i
      \pi /4} }{ 1- e^{
      \frac{i}{1-\Delta}
      (\pi-{\alpha}(\theta)/\sqrt{\xi})} },  
\label{3.3bb}
\end{eqnarray}
where
\begin{eqnarray}
{\tilde\alpha}(\theta) &:=& i(1-\Delta) 
 \sqrt{\xi} \left[1-\exp \left(i \frac{ {\alpha}(\theta) }
   {(1-\Delta) \sqrt{\xi}} \right) \right], 
\label{3.3b1}\\
\sigma(\theta) &:=& {\rm sign}( {\alpha}(\theta) ), 
\label{3.3c}
\end{eqnarray}
and
\begin{equation}
{\rm Erfc}(x):= \int_x^{+\infty} dt\, e^{-t^2}. 
\label{3.3d}
\end{equation}

We are mostly interested in the cases with $\Delta, \theta\ll 1$. 
Then, 
we have $\alpha(\theta)/\sqrt{\xi}\ll 1$, 
and therefore $\tilde\alpha(\theta)$ reduces to $\alpha(\theta)$. 
The second term in Eq.~(\ref{3.3bb}) is the contribution from the 
integral around the saddle point at $t=t_+$ along the contour $\tilde
C$. This term is not manifestly suppressed by $1/\sqrt{\xi}$. 
As far as $\alpha(\theta)$ is fixed, this term does not vanish 
in the limit $\xi\to \infty$. 
Of course, if we fix $\Delta$ and $\theta$ first, and take the 
limit $\xi\to \infty$, the argument of the error function goes to 
$+\infty$ and the function itself vanishes. 
However, 
$\alpha(\theta)$ vanishes at $\theta=\pi \Delta/(1-\Delta)$. 
Hence even for a very large value of $\xi$ there is always 
a region of $\theta$ in which this second term cannot be neglected. 
However, for $\theta$ in such a region, $\alpha(\theta)$ cannot 
be very large. 
Therefore, 
we can safely drop the second term in the exponent.  
On the other hand, the last term in Eq.~(\ref{3.3bb}), 
which is the contribution from the saddle point at $t=t_-$, 
is always suppressed by $1/\sqrt{\xi}$. Hence, this term does 
not give any significant contribution for $\xi\gg 1$. 
The first term in Eq.~(\ref{3.31})
can be dropped in the same manner for $\xi\gg 1$. 
Keeping only the terms which possibly remain 
in the limit $\xi\to \infty$, we finally obtain 
\begin{eqnarray}
\phi(\xi,\theta) &\approx &\exp \left( i\xi \cos \frac{
{\alpha}(\theta)}{\sqrt{\xi}} \right) \Theta
({\alpha}(\theta))\cr
&& -\frac{{\sigma}(\theta)}{\sqrt{\pi}} 
  e^{i\xi-\frac{i}{2}
   {\alpha}^2(\theta)}
 {\rm Erfc} 
  \left( \frac{ |{\alpha}(\theta)| }
    {\sqrt{2} } e^{-i
   \pi/4} \right) \cr
&& + (\theta\to -\theta).
\label{3.3b}
\end{eqnarray}

\begin{figure*}[t]
  \includegraphics[width=17.cm,clip]{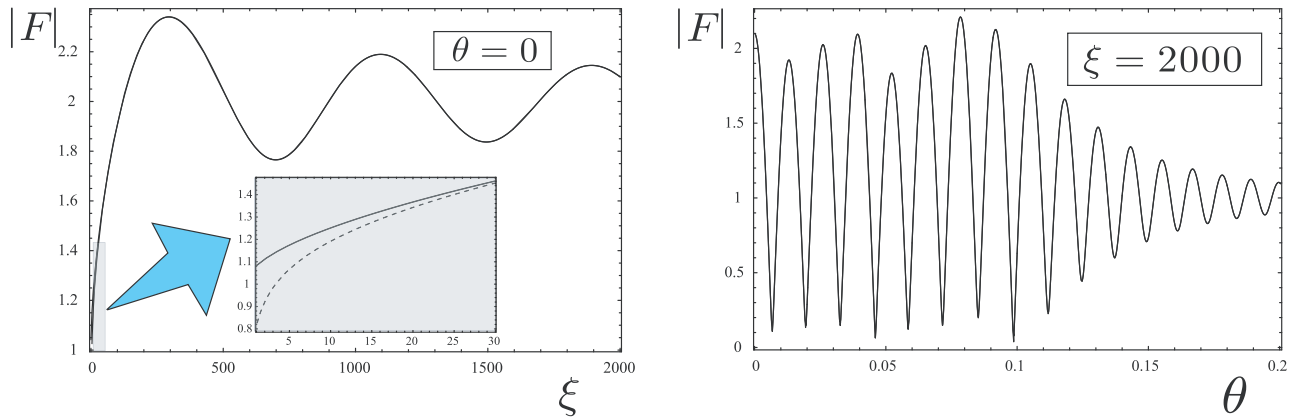}
\caption{
Comparison between the exact solution Eq.~(\ref{1.9}) and the
approximate one Eq.~(\ref{3.3bb}).
$\pi \Delta$ is $0.0025$.
Black line and dotted one correspond to the exact solution and
the approximate one, 
respectively.
We see that except for small $\xi$ the dotted line overlaps the
black one.
In the right panel,
the relative error is about $10^{-3}$.
}
\label{approx}
\end{figure*}

For illustrative purpose,
we compared the estimate given in Eq.~(\ref{3.3bb}) 
with the exact solution Eq.~(\ref{1.9})  
in Fig.~\ref{approx}. They agree quite well at $\xi\gg 1$. 
The deficit angle and the observer's direction 
are chosen to be $\Delta =0.0025$ and $\theta=0$, respectively.

\begin{figure}[h]
  \includegraphics[width=8.cm,clip]{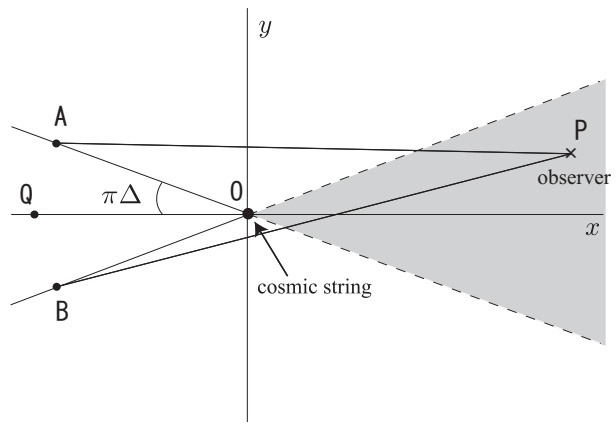}
\caption{
Configuration of the source, the cosmic string and the observer.
{\tt A} and {\tt B} are the positions of a source.
O and P are the positions of the cosmic string and the observer,
respectively.
In this figure,
the wedge {\tt AOB} is removed and thus {\tt A} and {\tt B} 
must be identified.
}
\label{configuration} 
\end{figure}

\subsection{Geometrical optics limit}
Geometrical optics limit corresponds to the limit 
$\xi \to \infty$ with $\Delta$ and $\theta$ fixed. 
In this limit $\alpha(\theta)$ also goes to $+\infty$, 
and hence the error function
in Eq.~(\ref{3.3b}) vanishes. 
Hence the waveform in the geometrical optics limit,
which we denote as ${\phi}_{\rm go}$, becomes 
\begin{eqnarray}
{\phi}_{\rm go}(\xi, \theta) & = &  
e^{i\xi \cos (\pi\Delta +\varphi) } \Theta (\pi\Delta +\varphi)
\cr
&& +
e^{i\xi \cos (\pi\Delta -\varphi) } 
\Theta (\pi\Delta-\varphi), 
 \label{go1}
\end{eqnarray}
where $\varphi$ is defined by Eq.~(\ref{1.25}).

Since $\phi$ and hence ${\phi}_{\rm go}$ are even in
$\theta$,
it is sufficient to consider the case with $\theta>0$.
In Fig.~\ref{configuration}, the configuration of 
the source, the lens and the observer is drawn in the 
coordinates in which the deficit angle $2\pi\Delta$ is manifest, i.e., 
the wedge {\tt AOB} is removed from the spacetime.  
Both points {\tt A} and {\tt B} indicate the location of the source. 
The lines {\tt OA} and {\tt OB} are to be identified. The angle made by 
these two lines is the deficit angle. 
The locations of the string and the observer are represented by 
{\tt O} and {\tt P}, respectively. 
In our current setup 
the distance between {\tt O} and {\tt A} 
($=r_o$) is taken to be infinite. 
When $\varphi > \pi \Delta$, only the source {\tt A} can be 
seen from the observer. 
This corresponds to the fact that 
only the first term remains for $\varphi>\pi \Delta$ 
in Eq.~(\ref{go1}). 
For $\varphi > \pi \Delta$, we have 
\begin{equation}
{\phi}_{\rm go}(\xi,\theta)=e^{i\xi \cos (\varphi+\pi\Delta)}. 
\label{go2}
\end{equation}
This is a plane wave whose traveling direction is 
$\varphi =-\pi \Delta$, which is the direction of 
$\overrightarrow{\tt AP}$ in 
Fig.~\ref{configuration} in the limit $r_o=|\overrightarrow{\tt AO}|
\to \infty$.  

For $|\varphi|< \pi\Delta$,
${\phi}_{\rm go}$ is
\begin{equation}
{\phi}_{\rm go}(\xi,\theta)=e^{i\xi \cos (\varphi-\pi\Delta)}
 +e^{i\xi \cos (\varphi+\pi \Delta)}. \label{go3}
\end{equation}
This is the superposition of two plane waves whose traveling 
directions are different by the deficit angle $2\pi\Delta$.
Hence amplification of the images and interference
occur for $|\varphi|< \pi\Delta$ as expected.

As we shall explain below, 
Eq.~(\ref{go1}) coincides with the one derived under 
the geometrical optics.
In geometrical optics,
wave form is given by \cite{T.Nakamura:1999}
\begin{equation}
\phi_{go} =
 \sum_{j}^{} {|u({\vec x_j})|}^{1/2} 
 \exp [i \omega T({\vec x_j}) -i \pi n_j], \label{go6}
\end{equation}
where $ {\vec x}$ represents a two-dimensional vector 
on the lens plane 
and $ T({\vec x}) $ represents the summation of time of flight of the 
light ray from the source to the point ${\vec x}$ on the lens plane 
and that from the point ${\vec x}$ to the observer. 
$ {\vec x_j} $ is a stationary point of $ T({\vec x}) $, and
$ n_j=0,1/2, 1 $ when $ {\vec x_j} $ is a minimum, saddle and 
maximum point of $ T({\vec x}) $, respectively. 
The amplitude ratio $ {|u({\vec x})|}^{1/2} $ is written as
\begin{equation}
u({\vec x}) = 1/ \det [ {\delta}_{a b}- 
{\partial}_a {\partial}_b \psi ({\vec x}) ], \label{go7}
\end{equation}
where $ \psi ({\vec x}) $ in Eq.~(\ref{go7}) is the deflection 
potential~\cite{Schneider} which is the integral of the gravitational 
potential of the lens along the trajectory between the source and the
observer.
Eq.~(\ref{go6}) represents that the wave form is
obtained by taking the sum of the amplitude ratio 
$ {|u({\vec x_j})|}^{1/2} $ of
each images with the phase factor 
$ e^{i \omega T({\vec x_j})-i\pi n_j} $.
If the lens is the straight string,
the spacetime is locally flat everywhere except for right on 
the string.
This means that the deflection potential $\psi({\vec x})$ 
is zero and hence the amplitude ratio is unity for 
all images~\cite{Schneider}
and the trajectory where the time of flight $T({\vec x})$
takes the extremal value is a geodesic in the conical
space, 
and $T({\vec x})$ of any geodesic takes minimum,
which means $n_j=0$.
There are two geodesics if the observer is in 
the shaded region in Fig.~\ref{configuration}.
The time of flight along the trajectory {\tt AP} is
\begin{equation}
T_{\tt A} =\lim_{r_o\to\infty} 
  |\overrightarrow{\tt AP}| 
\approx r_o+r\cos(\pi\Delta+\varphi), 
\label{go8}
\end{equation}
where $r \equiv |\overrightarrow{\tt OP}|$. 
The time of flight along the trajectory {\tt BP} is obtained by 
just replacing $\varphi$ with $-\varphi$. 
Hence, substituting (\ref{go8}) into (\ref{go6}),  
we find that the waveform in the geometrical optics 
is the same as Eq.~(\ref{go3}) except for an overall phase 
$e^{ir_o\xi}$. 
This factor has been already absorbed in the choice of the 
normalization factor $B$ in our formula (\ref{1.9}). 

We define the amplification factor 
\begin{equation}
F(\xi, \theta)=\frac{ {\phi} (\xi, \theta)}
 { {\phi}_{\rm UL} (\xi, \theta)}, \label{go4}
\end{equation}
where $ \phi_{\rm UL} $ is the unlensed waveform. 
Using Eq.~(\ref{go3}), the amplification 
factor of ${\phi}_{\rm go}$ for $|\varphi|<\pi \Delta$ is given by 
\begin{equation}
F_{{\rm go}}(\xi,\theta)
\approx 2 e^{-i \frac{\xi}{2} { (\pi \Delta)
 }^2} \cos (\pi\Delta \xi \varphi), \label{go5}
\end{equation}
where we have assumed $\varphi$ and $\Delta$ are small and dropped
terms higher than quadratic order. It might be more suggestive 
to rewrite the above formula into 
\begin{equation}
\left|F_{{\rm go}}(\xi,\theta)\right|
\approx 2 \cos (\pi\Delta \omega y),
\end{equation}
where $y= r\sin\varphi$. The distance from a node to the next 
of when the observer is moved in $y$-direction is 
$\lambda/\pi\Delta$,
where $\lambda$ is a wavelength. 
This oscillation is seen in the right panel of Fig.~\ref{approx}. 

\subsection{Quasi-geometrical optics approximation}
In the previous subsection,
we have derived the waveform in the limit
$\xi, |{\alpha}(\pm\theta)| \to \infty$ which corresponds to the
geometrical optics approximation.
Here we expand the waveform~(\ref{3.3b}) to the lowest order
in $1/{\alpha}(\pm\theta)$.
This includes the leading order corrections to the geometrical 
optics approximation due to the finite wavelength effects. 

For the same reason as we explained in the previous subsection,
we assume that $\Delta$ and $\varphi$ are small.
Using the asymptotic formula for the error function Eq.~(\ref{ap.5a}),
the leading order correction due to the finite wavelength, which 
we denote as $\delta{\phi}_{\rm qgo}$, is obtained as 
\begin{eqnarray}
\delta{\phi}_{\rm qgo}(\xi, \theta) 
 &=&-\frac{ e^{i\xi+i\pi/4} }{ \sqrt{2\pi}}
    \left({1\over \alpha(\theta)}+{1\over \alpha(-\theta)}
        \right) \cr
 &=&-\frac{ e^{i\xi+i\pi/4} }{ \sqrt{2\pi \xi} } 
    \frac{2\pi \Delta}{ (\pi \Delta)^2 -{\varphi}^2 },
\label{qgo1}
\end{eqnarray}
As is expected, the correction blows up for 
$|\varphi| \approx \pi\Delta$, where ${\alpha}(\theta)$ or 
${\alpha}(-\theta)$ vanishes, irrespectively of the value 
of $\xi$. 
In such cases, we have to 
evaluate the error function directly,  
going back to Eq.~(\ref{3.3b}). 

The expression on the first line in Eq.(\ref{qgo1}) manifestly 
depends only on $\alpha(\pm \theta)$ 
aside from the common 
phase factor $e^{i\xi}$. This feature remains true even if 
we consider a small value of $\alpha(\pm \theta)$.  
This can be seen by rewriting 
Eq.~(\ref{3.3b}) as 
\begin{eqnarray}
\phi(\xi,\theta) &\approx 
&\frac{e^{i\xi-\frac{i}{2}{\alpha}^2(\theta)}}{\sqrt{\pi}} 
{\rm Erfc} \left( \frac{ -{\alpha}(\theta)}{\sqrt{2i}} \right) + (\theta\to -\theta). \label{qgo1.5}
\end{eqnarray}
The common phase $e^{i\xi}$ 
does not affect the absolute magnitude of the wave.  
Except for this unimportant overall phase, 
the waveform is completely determined by $\alpha(\pm\theta)$. 

The geometrical meaning of these parameters $\alpha(\pm\theta)$ 
is the ratio of two length scales defined on the lens plane. 
To explain this, 
let us take the picture that a wave is composed of a superposition 
of waves which go through various points on the lens plane. 
In the geometrical optics limit the paths passing through 
stationary points of $T({\vec x})$, which we call the image points, 
contribute to the waveform. 
The first length scale is 
$r_s=|\alpha(\pm\theta)|/\sqrt{\xi}\times r$ 
which is defined as the separation between  
an image point and the string on the lens plane.  
In this picture we expect that 
paths whose pathlength is longer or shorter than the 
value at an image point by about one wavelength 
will not give a significant contribution because of 
the phase cancellation. Namely, only the paths which pass   
within a certain radius from an image point need 
to be taken into account. 
Then such a radius will be given by 
$r_F=\sqrt{\lambda r}$, which we call Fresnel radius. 
Namely, a wave with a finite wavelength can 
be recognized as an extended beam whose transverse size is 
given by $r_F$. 
The ratio of these two scales gives $\alpha(\pm\theta)$:
\begin{eqnarray*}
 |\alpha(\pm\theta)|={\sqrt{2\pi}r_s\over r_F}. 
\end{eqnarray*}
When $r_s\gg r_F$
, i.e., $\alpha(\pm\theta)\gg 1$, the 
beam width is smaller than the separation. 
In this case the beam image is not shadowed by the string, 
and therefore the geometrical optics becomes a good approximation. 
When $r_s\lesssim r_F$, i.e., 
\begin{equation}
 \alpha(\pm\theta)\lesssim 1, \label{qgo4}
\end{equation}
we cannot see the whole image of the beam, 
truncated at the location of the string. 
Then the diffraction effect becomes important.  
The ratio of the beam image eclipsed by the string 
determines the phase shift and the amplification of the wave 
coming from each image. 
If we substitute $|\varphi|\approx 0$ as a typical value, 
we obtain a rough criterion that the diffraction effect becomes 
important when
\begin{eqnarray}
\lambda \gtrsim 2\pi {(\pi \Delta)}^2 r, 
\end{eqnarray}
or $\xi \lesssim {(\pi \Delta)}^{-2}$ in terms of $\xi$. 

The same logic applies for a usual compact lens object. 
In this case the Fresnel radius does not change but the 
typical separation of the image from the lens is given by 
the Einstein radius $r_E\approx \sqrt{4GMr}$, where $M$ 
is the mass of the lens. Then the ratio between $r_E$ and 
$r_F$ is given by $r_E/r_F=\sqrt{GM/\lambda}$, which leads 
to the usual criterion that the diffraction effect becomes 
important when 
$\lambda\gtrsim GM$\cite{Ohanian,Bliokh,Bontz,Thorne,Deguchi}. 

From the above formula (\ref{qgo1}), 
we can read that the leading order corrections scales like 
$\propto\sqrt{\lambda/r}$. 
This dependencies on $\lambda$ and $r$ differ
from the cases that the lens is composed of a normal 
localized object, in which the leading order correction due to the finite
wavelength is ${\cal O}(\lambda /M)$~\cite{Takahashi:2004mc}.

The condition for the diffraction effect 
to be important~(\ref{qgo4}) can be also derived directly from
Eq.~(\ref{qgo1}).
In order that the current expansion is a good approximation,
${\phi}_{\rm qgo}$ must be smaller than ${\phi}_{\rm go}$. 
This requires that $1/\alpha(\pm\theta)\gg 1$, which is 
identical to (\ref{qgo4}).

\begin{figure}[t]
  \includegraphics[width=7.cm,clip]{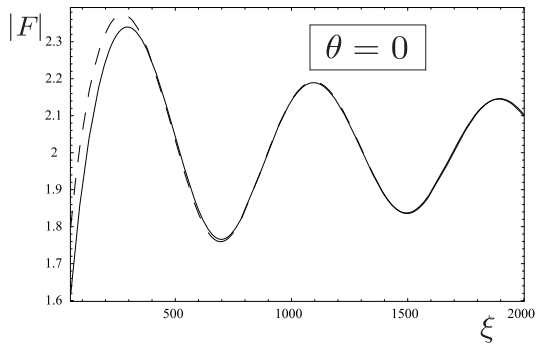}
\caption{
The absolute value of the amplification factor as a function of $\xi$
for $\theta=0$.
Black line and dashed one correspond to Eq.~(\ref{3.3b}) and the 
quasi-geometrical optics approximation,
respectively.
The string tension is chosen to be $G\mu=10^{-2}$.
}
\label{quasi1} 
\end{figure}

We plot the absolute value of the amplification factor under 
the quasi-geometrical optics approximation as dashed line 
in Fig.~\ref{quasi1}.
We find that the quasi-geometrical optics approximation is a good 
approximation for $\xi \gtrsim \Delta^{-2}$.
For $\xi \lesssim \Delta^{-2}$,
the quasi-geometrical optics approximation gives a 
larger amplification factor than the exact one.

In the quasi-geometrical optics approximation,
we find from Eqs.~(\ref{go3}) and (\ref{qgo1})
the absolute value of the amplification factor for $\varphi=0$ is
\begin{equation}
|F(\xi,0)| \approx 2 \left[ 1-\sqrt{ \frac{2}{\pi \xi {(\pi\Delta}^2}
 } \cos \left( \frac{\xi}{2} {(\pi\Delta)}^2+\frac{\pi}{4} \right) 
 \right]^{1/2}. \label{qgo5}
\end{equation}
From this expression,
we find that the position of the first peak of the amplification
factor lies at $\xi \approx 4.25 \times {(\pi\Delta)}^{-2}$, 
which can be also verified from Fig.~\ref{quasi1}.
For $\xi \lesssim \Delta^{-2}$ the present approximation 
is not valid, but we know that the amplification factor should 
converge to unity in the limit $\xi\to 0$, where $r_F$ is 
much larger than $r_s$. 

\begin{figure*}[t]
  \includegraphics[width=17.cm,clip]{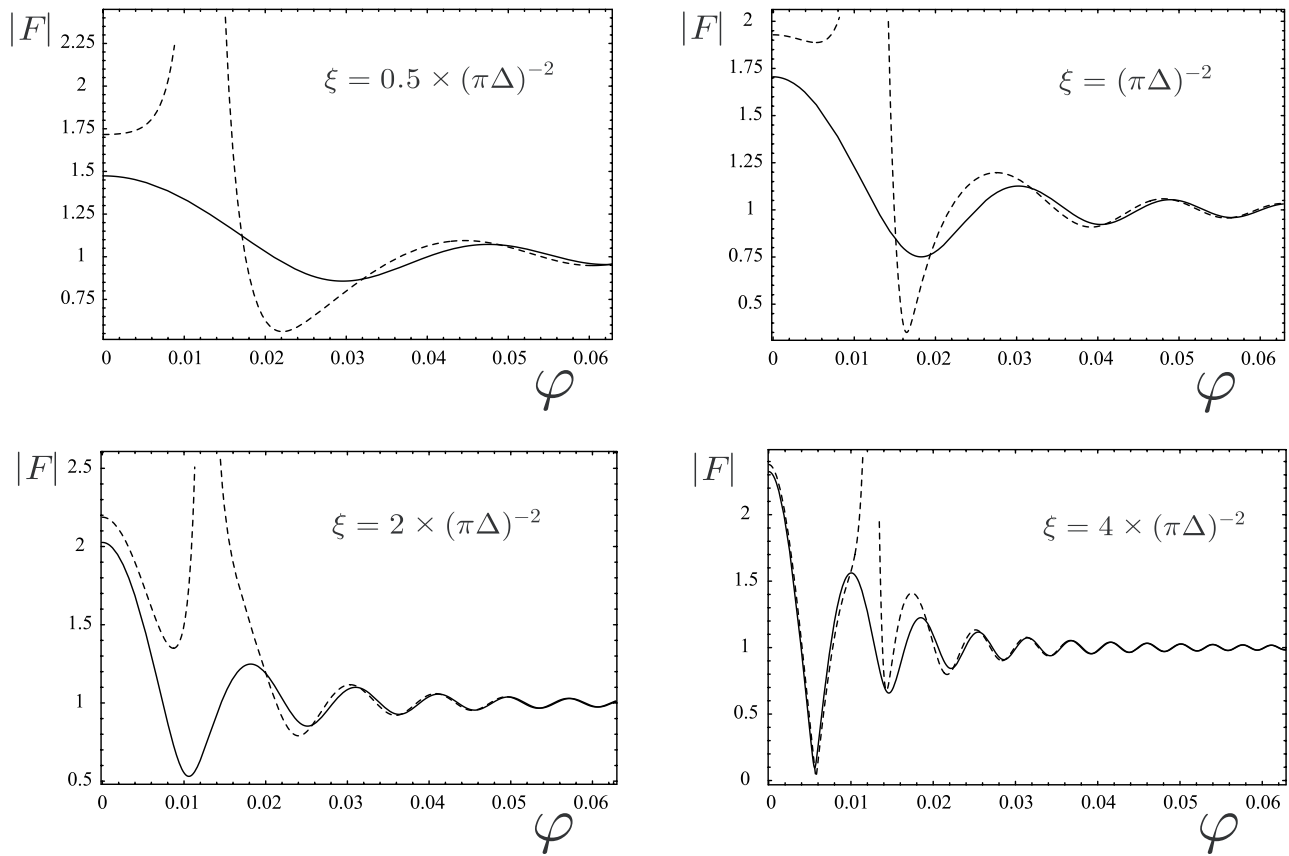}
\caption{
Black line and dotted one correspond to Eq.~(\ref{3.3b}) and the
quasi-geometrical optics approximation,
respectively.
The string tension is chosen to be $G\mu=10^{-3}$.
}
\label{quasi2} 
\end{figure*}

We show in Fig.~\ref{quasi2} the absolute value of the
amplification factor as a function of $\varphi$ for
four cases of $\xi$ around $\Delta^{-2}$.
Top left, top right, bottom left and bottom right panels
correspond to $\xi {(\pi \Delta)}^2=0.5,1,2$ and $4$,
respectively.
Black curves are plots for Eq.~(\ref{3.3b}) and the 
dotted ones are plots for the quasi-geometrical optics
approximation.
As is expected, the error 
of the quasi-geometrical optics approximation becomes 
very large near $\varphi=\pi\Delta$, 
where $\alpha(\theta)$ vanishes. 
As the value of $\xi$ increases, the angular region in which the
quasi-geometrical optics breaks down is reduced. 

Interestingly, 
the absolute value of the amplification factor deviates from unity
even for $ \varphi \gtrsim \pi \Delta$ which is not observed 
in the geometrical optics limit.
This is a consequence of diffraction of waves,
the amplitude of oscillation of the interference pattern 
becomes smaller as $\theta$ becomes larger,
which is a typical diffraction pattern formed 
when a wave passes through a single slit.
The broadening of the interference pattern due to the diffraction
effect means that the observers even in the region 
$|\varphi|> \pi \Delta$ can detect signatures of the 
presence of a cosmic string. 

But the deviation of the amplification from unity outside the wedge
$\varphi >\pi \Delta$ is rather small except for the special case
$\xi {(\pi \Delta)}^2 \approx 1$:
for $\xi {(\pi \Delta)}^2\ll 1$ the magnification is inefficient and
for $\xi {(\pi \Delta)}^2\gg 1$ the magnification itself does not occur.
Hence the increase of the event rates of lensing by cosmic strings
compared with the estimate under the geometrical optics approximation
could be important only when the relation $\xi {(\pi \Delta)}^2 \approx 1$
is satisfied.
If we take $D=10^{28} {\rm cm}$ and $\omega=10^{-3} {\rm Hz}$ which is in 
the frequency band of LISA(Laser Interferometer Space Antenna)\cite{lisa},
we find that the typical value of $G\mu$ is $\approx 2\times 10^{-9}$.

So far,
we have considered the stringy source rather than a point source.
Extension to a point source can be done in a similar manner to 
the case of the stringy source and is treated in Appendix B.
The result is 
\begin{eqnarray}
\phi(r,\theta,z)
   \approx -{1
   \over 4\pi D} 
     e^{i\omega D}
   {\cal F}\left( \frac{\omega rr_o}{D},\theta \right), \label{qgo7}
\end{eqnarray}
where $D=\sqrt{(r+r_o)^2+z^2}$ 
is the distance between the source and the observer. 
${\cal F}$,
which is defined by Eq.~(\ref{b1}),
is related to $\psi$ as
\begin{equation}
{\cal F}(x, \theta)= \big[ e^{-i\xi}\psi (\xi,\theta) \big] {\Big|}_{\xi \to x}+(\theta \to -\theta).
\end{equation}
Hence $\phi$ for the point source is similar to that
for the stringy source.
In particular,
assuming that $\Delta, \varphi \ll 1$, and keeping
terms which could remain for $\omega r, \omega r_o \gg 1$,
we have
\begin{eqnarray}
&F&\!\! \left(\frac{\omega rr_o}{D},\theta \right) \approx e^{-\frac{i}{2}\frac{\omega rr_o}{D} {(\pi \Delta-\varphi)}^2 } \cr
&&\times \frac{1}{\sqrt{\pi}} {\rm Erfc} \left( \frac{\varphi-\pi \Delta }{\sqrt{2i}} \sqrt{ \frac{\omega rr_o}{D} } \right)  
+(\theta \to -\theta). \label{qgo8}
\end{eqnarray}

\subsection{Simpler derivation of Eq.~(\ref{qgo1.5}).}
We have derived an approximate waveform~(\ref{qgo1.5})
which is valid in the wave zone from the exact solution of
the wave equation Eq.~(\ref{1.9}).
Here we show that Eq.~(\ref{qgo1.5}) can be obtained by a 
more intuitive and simpler method.
In the path integral formalism \cite{T.Nakamura:1999},
the wave form is given by the sum of the amplitude 
$\exp \left( i\omega T(s) \right)$ for all possible 
paths which connect the source and the observer.
Here $T(s)$ is the time of flight along the path $s$.
If the cosmic string resides between the source and the
observer,
the wave form will be given by the sum of two terms one 
of which is obtained by the path integral over the paths
which pass through the upper side of the string ($y>0$) in 
Fig.~\ref{configuration}, 
and the other through the lower side of it ($y<0$). 
The waveform coming from the former contribution 
will be given by
\begin{equation}
A \int_{-\infty}^{\infty} dz_{\tt Q}
  \int^{\infty}_0 dy_{\tt Q}\, e^{i\omega
 (|\overrightarrow{\tt AQ}|+|\overrightarrow{\tt QP}|)},  \label{sd1}
\end{equation}
where ${\tt Q}=(0,y_{\tt Q},z_{\tt Q})$ is a point on the lens plane specified by 
$x=0$.  
One can determine the normalization constant $A$ by a little more 
detailed analysis, but we do not pursue it further here.  
By integrating Eq.~(\ref{sd1}),
we recover the first term in Eq.~(\ref{qgo1.5}). 


\subsection{Long wavelength limit}
For completeness, we consider the case in which 
the wavelength is longer than the distance from the 
string $\xi \lesssim 1$. 
In this limit, 
the first few terms in Eq.~(\ref{1.9}) dominate, 
and we find 
\begin{equation}
\phi (\xi,\theta) \approx \frac{1}{1-\Delta}+i\frac{e^{-\Delta \log 2-\pi \Delta/2 i}}{\Gamma (1+\Delta)} {\xi}^{1+\Delta} \cos \theta. \label{3.17}
\end{equation}
In particular, 
for $\xi \to 0$ Eq.~(\ref{3.17}) becomes
${(1-\Delta)}^{-1}$ which is larger than unity.
This differs from the cases of gravitational lensing by a normal
compact object, where the amplification becomes unity in the long
wavelength limit. 
The reason why the amplification differs from unity even in
the long wavelength limit is that the space has a deficit angle 
and hence the structure at the spatial infinity 
is different from the usual Euclidean space. 
Waves with very long wavelengths do not feel the local 
structure of string. 
However, uniform amplification of waves should occur 
as a result of total energy flux conservation because 
the area of the asymptotic region at a constant distance 
from the source is reduced due to the deficit angle. 
In this sense such modes feel the existence of a string. 

\section{Connections to observations}
\subsection{Compact binary as a source}
In this section, 
we consider compact binaries as sources of gravitational waves.
Gravitational waves from compact binaries are clean in the
sense that the waves are almost monochromatic:
the time scale for the frequency to change 
is much longer than the orbital period of the binary 
except for the phase just before plunge.
Hence interference between two waves coming from both sides
of the cosmic string could be observed by future
detectors.

Since each compact binary has a finite lifetime,
lensing events can be classified roughly into two cases. 
If the difference between the times of flight along two geodesics 
is larger than the lifetime of the binary,
we will observe two independent waves separately 
at different times.
On the other hand,
if the time delay is shorter than the lifetime,
what we observe is the superposition of two waves. 

The remaining lifetime of the binary $T_{\rm life}$ 
when the period of the gravitational waves
measured by an observer is $P_{GW}$ is estimated as 
\begin{equation}
T_{\rm life} \approx 9.2 \times 10^{-4} \frac{1}{ {(1+z_S)}^{5/3} }
 \frac{ {(1+\eta)}^{1/3} }{\eta} { \left( \frac{P_{GW}}{GM} \right)
 }^{5/3} P_{GW}, \label{com1}
\end{equation}
where $\eta$ is the mass ratio of the binary ($\eta \le 1$), 
$M$ is the mass of the more massive star in the binary 
and $z_S$ is the source redshift.

\begin{figure*}[t]
  \includegraphics[width=17.cm,clip]{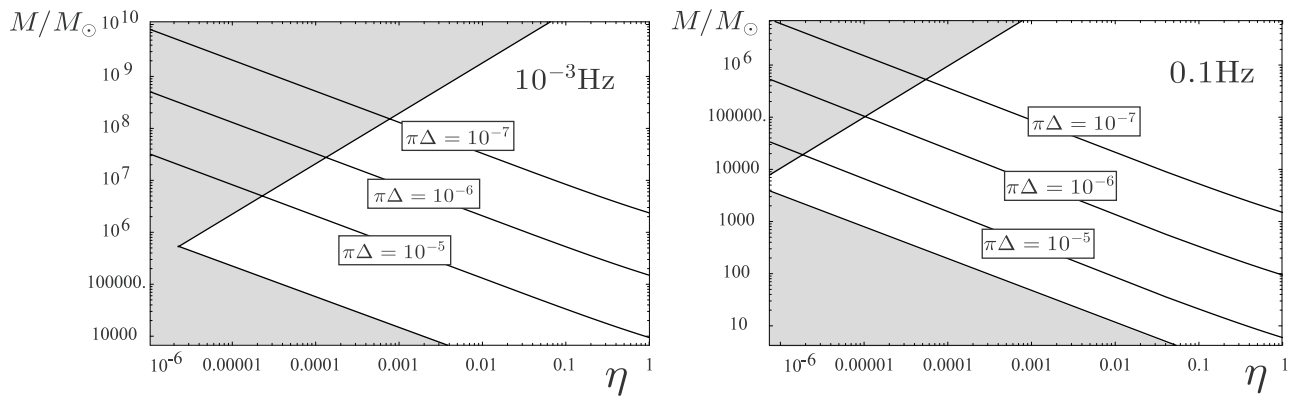}
\caption{
Plots of regions where Eq.~(\ref{com3}) is satisfied for three different
values of the string parameter.
Left and right panels are for $10^{-3} {\rm Hz}$ and $0.1 {\rm Hz}$
which are the frequency bands LISA and DECIGO have best sensitivities.
Shaded regions are plotted under the assumptions that the signals satisfy $SN>10$,
the redshift of the source is $1$ and $3$ year observations.
}
\label{lisa-decigo} 
\end{figure*}

The time delay $T_{\rm delay}$ is 
\begin{equation}
T_{\rm delay} \approx 2 \frac{r r_o}{D} 
  \varphi \pi \Delta. \label{com2}
\end{equation}
Taking the typical values of parameters as
$r r_o/D  =1 {\rm Gpc}$ and 
$\varphi=\pi \Delta$, 
the condition $T_{\rm life} \gg T_{\rm delay}$ gives the upper bound 
on the mass $M$, 
\begin{equation}
M \ll 8\times 10^3 \frac{ {(1+\eta)}^{1/5} }{ {\eta}^{3/5} } { \left( \frac{\pi \Delta}{10^{-5}} \right) }^{-6/5} { \left( \frac{P_{GW}}{10^3 {\rm sec}} \right) }^{8/5} M_{\odot}. \label{com3}
\end{equation}
The time scale for the orbital 
frequency of the binary to change is the same 
order as $T_{\rm life}$.
Hence the condition $T_{\rm life} \gg T_{\rm delay}$ implies
that the frequencies of two waves are almost the same.
The left and right panels in Fig.~\ref{lisa-decigo} which
correspond to different frequencies of gravitational waves
show the region where the condition Eq.~(\ref{com3}) is 
satisfied for three different values of string parameter $\Delta$.
The shaded area represent the parameter region beyond the
detector's sensitivities. 
In the left and right panels 
we assumed, respectively, that the threshold value for detection 
in strain amplitude for LISA and DECIGO(DECihertz Interferometer Gravitational
wave Observatory)\cite{skn01}/BBO(Big Bang Observer)\cite{bbo}, which 
are given by $10^{-20}{\rm Hz}^{-1/2}$ and $10^{-23}{\rm Hz}^{-1/2}$. 
We find that both cases $T_{\rm life} \gg T_{\rm delay}$ and 
$T_{\rm life} \ll T_{\rm delay}$ can occur both for LISA and
BBO/DECIGO. 

\subsection{Waveform}
We can easily extend our waveform~(\ref{1.9}) to the case that
the frequency of the source changes in time.
Let us write the source as $\frac{1}{1-\Delta} S(t) \delta(r-r_o) \delta
(\theta-\pi) \delta (z)$. 
The Fourier transformation of $S(t)$ is defined by 
\begin{equation}
S(t) =\int_{-\infty}^{\infty} d\omega ~e^{-i \omega t}S_{\omega}. \label{com4}
\end{equation}
Denoting the solution $\phi(t,\vec{x})$ for a monochromatic source 
obtained in the previous sections by ${\phi}_{\omega} (\vec{x})$,
$\phi$ can be written as 
\begin{equation}
\phi (t, \vec{x})= \int_0^{\infty} d\omega ~e^{-i\omega t} S_{\omega} {\phi}_{\omega}(\vec{x})+c.c.,
\end{equation}
where we assumed that $S(t)$ is real.
Substituting Eq.~(\ref{qgo7}) to the above expression,
we have
\begin{eqnarray}
\phi (t,\vec{x}) \approx -\frac{1}{4\pi D} \int_0^{\infty} d\omega ~e^{-i\omega (t-D)} {\cal F} \left( \frac{\omega rr_o}{D},\theta \right) S_\omega \nonumber \\
 +c.c. \label{com4.1}
\end{eqnarray}

Eq.~(\ref{com4.1}) is a general formula which applies to 
any time dependent source.
Here we consider the special case in which $S(t)$ takes the
form 
\begin{equation}
S(t)=\cos \left( \int_0^t dt' ~\Omega(t') \right),
\end{equation}
with 
\begin{equation}\Omega(t)=\omega_o+{\dot \omega_o}t, 
\end{equation}
where ${\dot \omega_o}/{\omega_o}^2 \ll 1$ and $\omega_o >0$ 
are assumed. This represents a quasi-monochromatic source with 
its frequency slowly changing. 
Then $S_\omega$ is
\begin{equation}
S_\omega =\frac{1}{\sqrt{2\pi {\dot \omega_o}} } \left( e^{-i \frac{ {(\omega+\omega_o)}^2 }{2{\dot \omega_o}} +i\pi /4}+e^{i \frac{ {(\omega-\omega_o)}^2 }{2{\dot \omega_o}} -i\pi /4} \right). \label{com4.3}
\end{equation}
Substituting Eqs.~(\ref{qgo8}) and (\ref{com4.3}) 
into Eq.~(\ref{com4.1}),
and using the method of the steepest descent, we have
\begin{eqnarray}
\phi (t, \vec{x}) &\approx& \frac{1}{4\pi\sqrt{\pi} D} 
{\rm Erfc} \left( \frac{ \varphi-\pi \Delta}{\sqrt{2i}} \sqrt{
	    \frac{\Omega ( T(\varphi) ) rr_o}{D} } \right) \cr
 &&
   \times e^{-iT(\varphi)
(\omega_o+\frac{1}{2} {\dot \omega_o}T(\varphi))} 
+c.c.\cr
&&\qquad 
+(\varphi \to -\varphi), \label{com4.4} 
\end{eqnarray}
with
\begin{eqnarray}
T(\varphi) = t-D+\frac{rr_o}{2D} {(\pi \Delta-\varphi)}^2. \label{com4.5}
\end{eqnarray}
This represents a superposition of two waves coming from both
sides of the string whose arrival times differ by 
$|T(\varphi)-T(-\varphi)| =\frac{2rr_o}{D} \pi \Delta |\varphi|$. 
In the preceding subsections, 
we study the waveforms observed in two
cases with  
$T_{\rm life}\gg T_{\rm delay}$ and $T_{\rm life}\ll T_{\rm delay}$. 

\subsubsection{$T_{\rm life} \gg T_{\rm delay}$}
As we have explained in the preceding subsection,
what we observe is a superposition of two waves
in this case.
Because the relative phase difference of these waves 
slowly increases or decreases in time due to the 
frequency change of the binary source and the 
optical path difference between two geodesics,
we will observe the beat if the amplitude of the
integrated relative phase difference over observation 
time is larger than ${\cal O}(1)$.

The condition that the beat is observed can be derived
as follows.
If we denote the total observation period by $T_{\rm obs}$,
then from Eq.~(\ref{com4.4}) the integrated relative phase difference 
is $2\pi \Delta \varphi D {\dot \omega_o} T_{\rm obs}$, 
where both $r$ and $r_o$ are assumed to be $O(D)$. 
Hence we can observe the beat if
\begin{equation}
T_{\rm obs} \gtrsim \frac{1}{2\pi \Delta \varphi 
D {\dot \omega_o}}. \label{com10}
\end{equation}

Because $T_{\rm life}$ is roughly the same as the time 
scale for the frequency of the binary to change, i.e.
$T_{\rm life} \sim {\omega_o}/{\dot \omega_o}$,
Eq.~(\ref{com10}) can be written as
\begin{equation}
T_{\rm obs} \gtrsim \frac{T_{\rm life}}
    {2\pi \Delta \varphi D\omega_o}. \label{com12}
\end{equation}

\begin{figure}[t]
\includegraphics[width=8cm,clip]{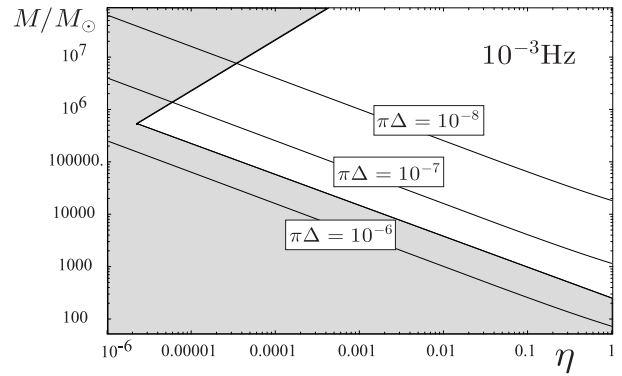}
\caption{Plot of the region where Eq.~(\ref{com13}) is satisfied.
The frequency of the gravitational waves is assumed to be
$10^{-3} {\rm Hz}$.}
\label{lisa2} 
\end{figure}

If $T_{\rm obs}$ is fixed, 
e.g. $T_{\rm obs}\sim 3 {\rm yr}$ for LISA, 
Eq.~(\ref{com12}) is written as an lower bound on $M$.
For $T_{\rm obs} =3 {\rm yr}$ 
and $P_{GW}=10^3 {\rm sec}$,
Eq.~(\ref{com12}) becomes 
\begin{equation}
M \gtrsim 2.6 \times \frac{ {(1+\eta)}^{1/5} }{ {\eta}^{3/5} } { \left( \frac{\pi \Delta}{10^{-5}} \right) }^{-6/5} {\left( \frac{P_{GW}}{10^3 {\rm sec}} \right) }^{13/5} M_{\odot}. \label{com13}
\end{equation}
We show in Fig.~\ref{lisa2} the region where Eq.~(\ref{com13}) 
is satisfied for LISA with $T_{\rm obs} =3 {\rm yr}$. 
We find that if $G\mu \lesssim 2.8 \times 10^{-8}$ which is about
one order of magnitude below the current upper bound,
LISA will detect the beat of gravitational waves for all 
observable ranges in $(\mu,M)$ space as long as 
$T_{\rm life}\gg T_{\rm dely}$
\footnote{
Since the lensing probability is not expected to be high, 
we need a large number of events to detect a lensing event. 
In such a situation, what gravitational wave detectors can detect 
is a superposition of various waves. Hence, signal will almost 
always have beat even if we ignore the lensing effect.}.

\subsubsection{$T_{\rm life} \ll T_{\rm delay}$}
\begin{figure*}[t]
  \includegraphics[width=17.cm,clip]{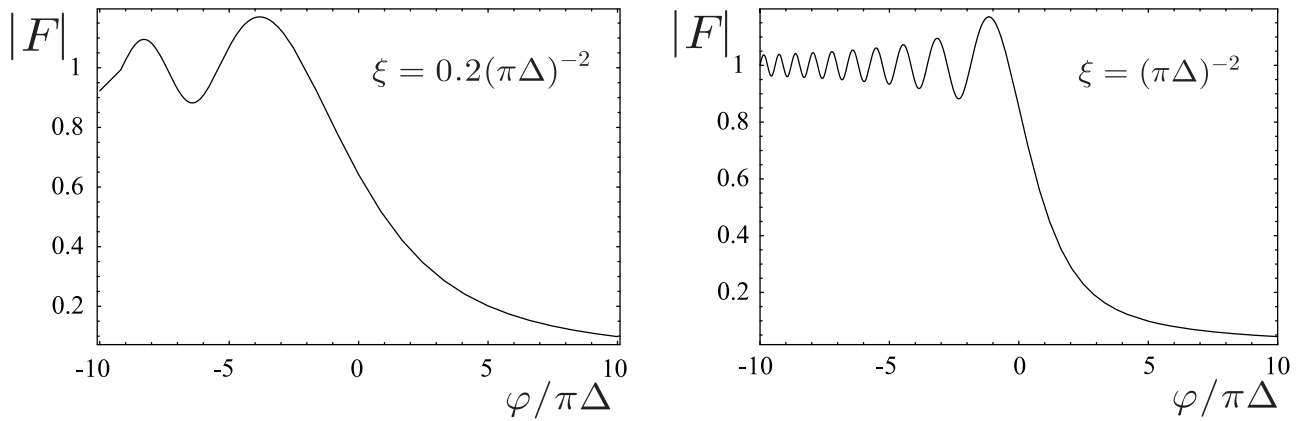}
\caption{
The absolute value of the amplification factor for
$T_{\rm life} \ll T_{\rm delay}$ as a function of $\varphi$.
Left and right panels correspond to 
$\xi=0.2 {(\pi \Delta)}^{-2}$ and ${(\pi \Delta)}^{-2}$,
respectively.
}
\label{oneside} 
\end{figure*}

If $T_{\rm delay} \gg T_{\rm life}$,
we observe the waveform of either the first term or the second one in
Eq.~(\ref{com4.4}) at a given time.
We show in fig.~\ref{oneside} the amplification of the wave corresponding
to the first term in Eq.~(\ref{com4.4} ) as a function of 
$\varphi-\Delta\pi$ 
normalized by $1/\sqrt{\omega rr_o/D}$, 
which is nothing 
but $-\alpha(\theta)$ in the case discussed 
in Sec.\ref{sec:behaviorsofsolution}.
We find that the amplification 
approaches zero more slowly for 
$\varphi -\pi \Delta>0$ and oscillates around
unity for $\varphi -\pi\Delta <0$ and 
the angular size in which non-trivial oscillations due to 
the diffraction effect
can be observed is given by $1/\sqrt{\omega rr_o/D}$. 
Since 
$T_{\rm delay}\approx (rr_o/D) \varphi\pi\Delta<T_{\rm life}\ll
\omega^{-1}$ in the present case, we have $(rr_o/D)
 (\pi\Delta)^2\agt (rr_o/D)\pi\Delta\varphi\gg 1$. 
Therefore this angular size of oscillation is much smaller
than $\pi\Delta$. Hence it will be very difficult to detect 
a lensing event in which this diffraction effect is relevant. 

\subsection{Estimation of the event rate}
In this section,
we estimate the detection rate of the gravitational 
lensing caused by cosmic strings for planned gravitational 
wave detectors such as LISA, DECIGO and BBO.

It is well known that string network obeys the scaling solution
where the appearance of the string network at any time looks 
alike if it is scaled by the horizon size. 
There are a few dozen strings spread crossing the horizon volume and a 
number of string
loops~\cite{Albrecht:1989mk,Bennett:1989yp,Allen:1990tv}.
Since the horizon scale increases in the comoving coordinates as 
time goes, the number of strings increase if there is no 
interaction between them. However, since strings are typically 
moving at a relativistic speed, they frequently intersect with 
each other. As a result reconnection between strings occurs, 
reducing the number of long strings which extend over 
the horizon scale. During the process of 
reduction of the number of long strings 
a large number of string loops are formed, but 
they shrink and decay via gravitational radiation. 
Due to the balance of two effects, 
the number of long strings in a horizon volume 
remains almost constant in time.

The reconnection probability $p$ is essentially $1$ 
for gauge theory solitons~\cite{Matzner}
because reconnection allows the flux inside the string to
take an energetically favorable shortcut.
For F-strings, the reconnection is a quantum process and 
its probability is roughly estimated as 
$ p\sim g^2_s $, where $ g_s $ is the string coupling and
is predicted in \cite{Jackson:2004zg} that 
\begin{equation}
10^{-3} \lesssim p \lesssim 1. \label{es1}
\end{equation}
For D-strings, the reconnection probability might be 
$ 0.1 \lesssim p \lesssim 1 $ \cite{Jackson:2004zg}.
If the reconnection probability is less than $1$,
the number of long strings is expected to 
be $ p^{-1} $ times larger than that in the case with $p=1$.
Therefore it is expected that in the context of cosmic strings 
motivated by superstring theory 
the number of long strings in a horizon volume 
can be $ 10^3 $ or more.

To estimate the event rate for the gravitational lensing, 
here we consider a compact binary 
(such as binary neutron stars and/or black holes)
as a source of gravitational waves.
There are large uncertainties about the event rate of MBH (massive black
hole) merger detected by LISA or DECIGO/BBO.
Several authors \cite{Haehnelt:1994wt,Islam:2003nt,Ioka:2005pm} 
employed a model in which MBH mergers are associated with the 
mergers of host dark matter halos to estimate the event rate 
of MBH-MBH mergers.
In this model,
the event rate is dominated by halos with the minimum mass 
$M_{min}$ above which halos have a central MBH and 
some scenario predict that the event rate could reach 
$\sim 10^4 $ events/yr.
For DECIGO/BBO, the binary neuron stars will be observed
$\sim 10^5$ events/yr.

The probability of lensing for a single source by an infinite
straight cosmic string both at cosmological distances is 
\begin{equation}
P \simeq 3\times 10^{-6} \left( \frac{\pi \Delta}{10^{-5}} \right). \label{es3}
\end{equation}
Eq.~(\ref{es3}) is derived under the geometrical optics
approximation.
In section III,
we found that the signal of lensing by cosmic strings 
(the interference pattern of gravitational waves at 
detectors) extends over an angular scales 
larger than the deficit angle $2\pi \Delta$
when the diffraction effect is marginally important.
This is a well known fact for the gravitational lensing by 
usual stellar objects \cite{Takahashi:2003ix,Ruffa:1999}.
As we estimated in section III,
the critical distance $D_c$ below which the diffraction effect
becomes important is
\begin{equation}
D_c=50 \left( \frac{P_{GW}}{10^{3} {\rm sec}} \right) { \left( \frac{\pi
	\Delta}{10^{-5}} \right) }^{-2} {\rm kpc}.  
\label{es4}
\end{equation}
Therefore the probability of lensing by cosmic strings may 
be enhanced due to the diffraction effect 
for $ \pi\Delta \approx 10^{-7} $ at LISA band ($P_{GW}\approx 10^3$sec)
and for $ \pi\Delta \approx 10^{-8} $ at DECIGO/BBO band 
($P_{GW}\approx 10 $sec). 

Assuming the prospective values of the parameters that determines
the rate of lensing events $\dot n$, we obtain 
\begin{equation}
\dot{n} \sim 3 f { \left( \frac{p}{0.1} \right) }^{-1} \left( \frac{\pi
		     \Delta}{10^{-5}} \right) \left(
		     \frac{\dot{n}_S}{10^5 {\rm yr}^{-1}} \right) {\rm
		     yr}^{-1}, \label{es5} 
\end{equation}
where $f$($>1$) denotes the numerical factor arising from 
the enhancement of the lensing probability due to the diffraction effect.
$\dot{n}_S=10^5$ is almost upper bound on the total 
event rate of neutron star mergers detectable by DECIGO/BBO.
If the event rate is even higher, the number of events 
becomes comparable to or larger than the number of frequency bins.  
Then we will not be able to distinguish each event, and 
undistinguishable signals become confusion noise. 
In the case of LISA, this bound on $\dot n_S$ is even lower. 
Unfortunately, 
a large number of lensing events by cosmic strings 
can be expected only for marginally large $\pi\Delta(\approx G\mu/4)$ 
with a small reconnection probability $p$. 

Finally we briefly comment on the validity of the assumption that 
most of cosmic strings can be treated as straight ones 
in studying gravitational lensing by them. 
In geometrical optics approximation,
only light paths which satisfy the Fermat's principle contribute to the
amplification factor.
If we take into account the finiteness of the wavelength,
the trajectories whose optical path differences are less than 
a few times of its wavelength will dominantly 
contribute to the amplification factor.
In terms of the distance on the lens plane
($x=0$-plane in Fig.~\ref{configuration}), 
the optical paths within $\alt\sqrt{\lambda D}$
from the intersection of the geodesic 
will give a dominant part of the amplification factor. 

In the standard literature,
the typical size of small-scale structure of a long string 
is given by the
gravitational back-reaction scale $\sim 50 G\mu t$,
where $t$ is a cosmic time \cite{Vilenkin:2005jg}.
But this is not an established argument and some recent studies
suggest that the smallest size of the wiggles could be much smaller
than $50 G\mu t$ \cite{Siemens:2001dx,Siemens:2002dj}.
If we assume here that the smallest size of the wiggles is
$50 G\mu t$,
then the condition that the straight string approximation is good
is $\sqrt{\lambda D} \lesssim 50 G\mu t$.
Substituting the appropriate values of the parameters,
it gives the condition,
\begin{eqnarray}
&&1\gtrsim \frac{ \sqrt{\lambda D}}{50 G\mu t} \nonumber \\
&&=8 \times 10^{-5} { \left( \frac{\pi \Delta}{10^{-5}} \right) }^{-1} { \left( \frac{\lambda}{10^{13} {\rm cm}} \right) }^{1/2} { \left( \frac{D}{10^{26} {\rm cm}} \right) }^{1/2}. \label{es6}
\end{eqnarray}
Hence approximating a cosmic string by a straight one is good for
wide range of possible values of the parameters.

\section{Summary}
We have constructed a solution of the Klein-Gordon equation for a 
massless scalar field in the flat spacetime with a deficit
angle $2\pi\Delta\approx 8\pi G\mu$ caused by 
an infinite straight cosmic string.
We showed analytically that the solution in the short wavelength
limit reduces to the geometrical optics limit. 
We have also derived the correction 
to the amplification factor obtained 
in the geometrical optics approximation
due to the finite wavelength effect 
and the expression in the long wavelength limit.

The waveform is characterized by a ratio of two different length scales.
One length scale $r_s$ is defined as the separation between  
the image position on the lens plane in the geometrical optics 
and the string. 
We have two $r_s$ since there are two images 
corresponding to which side of the string the ray travels. 
(When the image cannot be seen directly, we assign a negative 
number to $r_s$.)
The other length scale $r_F$, which is called Fresnel radius, is 
the geometrical mean of the wavelength and the typical separation 
among the source, the lens and the observer. 
The waveform is characterized by the ratios between $r_s$ and 
$r_F$. 
If $r_F>r_s$, 
the diffraction effect becomes important and the interference
patterns are formed. 
Even when the image in the geometrical optics is not directly 
seen by the observer, the interference patterns remain.  
In contrast, 
in the geometrical optics
magnification and interference occur 
only when the observer can see two images which travel 
both sides of the string.
Namely, the angular range where lensing signals exist  
is broadened by the diffraction effect. 
This broadening may increase the lensing probability 
by an order of magnitude compared with that 
estimated by using the geometrical optics when the 
distance to the source is around the critical distance 
$D_c$ given in Eq.~(\ref{es4}).

We finally estimated the rate of lensing events 
which can be detected by LISA and DECIGO/BBO
assuming BH-BH or NS-NS mergers as a source of gravitational waves. 
For possible values of the parameters that determines the event
rate such as string reconnection rate, string tension and the event rate 
of the unlensed mergers,
the lensing event rate could reach several per yr.

\acknowledgements

T.S. thanks Kunihito Ioka, Takashi Nakamura and Hiroyuki Tashiro 
for useful comments. 
This work is supported in part 
by Grant-in-Aid for Scientific Research, Nos.
14047212 and 16740141, 
and by that for the 21st Century COE
"Center for Diversity and Universality in Physics" at 
Kyoto university, both from the Ministry of
Education, Culture, Sports, Science and Technology of Japan.

\appendix
\section{Derivation of Eq.~(\ref{3.3b})}
Here we derive a formula Eq.~(\ref{3.3b}) from the
integral representation of the solution Eqs.~(\ref{3.31}) 
and (\ref{3.3}).
As we explained in the sec.III,
we have to calculate the integral for $\alpha (\theta)>0$ and
$\alpha (\theta)<0$ separately.

\subsubsection{$\alpha (\theta)<0$}
In this case,
there is no contributions from the pole $t_{\ast}$.
Since $\xi \gg 1$,
the integral 
\begin{equation}
\psi (\xi,\theta)=\frac{1}{1-\Delta} \frac{1}{2i\pi} \int_C dt~
\frac{e^{\xi \sinh t}}{1-e^{-\frac{1}{1-\Delta}(t-i\frac{\pi}{2}+i\alpha (\theta) /\sqrt{\xi}  +\epsilon)}}, \label{ap.1}
\end{equation}
is dominated from the two regions
$ |t \pm i\frac{\pi}{2}| \lesssim 1/\sqrt{\xi}$,
where $\pm i\frac{\pi}{2}$ is the saddle points of
$e^{\xi \sinh t}$.

Let us first calculate the integral around $i \pi/2$.
We cannot apply the method of steepest descent where the denominator 
of the integrated function is replaced with the value at
$t=i\pi/2$ because the pole $t_{\ast}$ of the integrand
can lie in the region $|t_{\ast}-i\pi/2| \lesssim 1/\sqrt{\xi}$
and the denominator is no longer constant around $i \pi/2$.

Fortunately the integral can be approximated written by the special
function which can be evaluated easily.
We first do the transformation of variable such that
$t-i\pi/2=e^{i\pi/4}u$ (u:real number) which corresponds 
to the deformation of the contour of the integral from $C$ 
to ${\tilde C}$ as shown in Fig.~\ref{contour}.
Expanding $e^{\xi \sinh t}$ around $i \pi/2$ to second 
order in $u$ and the denominator of the integral to the 
first order in $u$ gives the integral
\begin{equation}
\frac{1}{2i\pi}\exp \left( i\xi +\frac{i \alpha (\theta)}{ (1-\Delta) \sqrt{\xi} } \right) \int_{-\infty}^{\infty} du \frac{ e^{-\frac{u^2}{2}} }{u+ e^{i\pi/4} {\tilde \alpha}(\theta)}, \label{ap.2}
\end{equation}
where ${\tilde \alpha}(\theta)$ is defined by Eq.~(\ref{3.3b1}).
Hence we need to evaluate the integral
\begin{equation}
I(x) := \int_{-\infty}^{\infty} du \frac{ e^{- \frac{u^2}{2}} }{u-x}
=\int_{\infty}^{\infty} du \frac{ e^{-\frac{ {(u+x)}^2 }{2}} }
{u-i\epsilon}, \label{ap.3}
\end{equation}
where $\epsilon$ was introduced to remember that the imaginary 
part of $x=-e^{i\pi/4}\tilde\alpha(\theta)$ 
is positive when $\tilde\alpha(\theta) <0$. 
This integral is given by an error function as
\begin{equation}
I(x)=2i\sqrt{\pi} e^{-{x}^2/2} {\rm Erfc} \left( -i \frac{x}{2} \right) \equiv 2i\sqrt{\pi} e^{-{x}^2/2} \int_{-i \frac{x}{2}}^{\infty} dt e^{-t^2}. \label{ap.4}
\end{equation}
This can be derived by solving a differential equation 
\begin{equation}
\frac{d}{dx} I(x) = -\sqrt{2\pi}-x I(x), \label{ap.5}
\end{equation}
which follows from the definition of $I(x)$, 
with the boundary condition that $I(0)=i\pi$.

Asymptotic formulas for the error function are 
\begin{eqnarray}
 {\rm Erfc}(z) & = & e^{-z^2}\left({1\over 2z}-{1\over 4z^3}+\cdots\right),
\cr &&
\qquad ({\rm for}~-{\pi\over 4}<{\rm arg}\, z
  <{\pi\over 4}~{\rm and}~
   |z|\to\infty),
\cr
         & = & \sqrt{\pi}+
 e^{-z^2}\left({1\over 2z}-{1\over 4z^3}+\cdots\right),
\cr &&
 \qquad ({\rm for}~{3\pi\over 4}<{\rm arg}\, z
  <{5\pi\over 4}~{\rm and}~ |z|\to\infty),
\cr
         & = & \sqrt{\pi}+
 {\sqrt{\pi}\over 2}-z+{1\over 3}z^3+\cdots, 
\cr &&
 \qquad ({\rm for}~|z|\ll 1). \label{ap.5a}
\end{eqnarray}

Using Eq.~(\ref{ap.3}),
the integral Eq.~(\ref{ap.2}) becomes
\begin{equation}
\frac{1}{\sqrt{\pi}} \exp \left( i\xi+\frac{i \alpha (\theta)}{ (1-\Delta) \sqrt{\xi} }-\frac{i}{2} {\tilde \alpha}^2 (\theta) \right) {\rm Erfc} 
\left( \frac{ {\tilde \alpha}(\theta) }{\sqrt{2}} e^{3i\pi/4} \right). \label{ap.6}
\end{equation}

Next let us calculate the integral Eq.~(\ref{ap.1}) around $-i \pi/2$.
Since the pole $t_{\ast}$ is far from $-i\pi/2$,
we can approximate the denominator of the integrated function as a 
constant and apply the usual saddle point method.
This gives
\begin{equation}
\frac{1}{\sqrt{2\pi \xi}} \frac{1}{1-\Delta} \frac{ e^{-i\xi +i\pi/4} }{1- e^{ \frac{i}{1-\Delta} (\pi-\alpha (\theta)/\sqrt{\xi} ) }}. \label{ap.7}
\end{equation}
The sum of Eqs.~(\ref{ap.6}) and (\ref{ap.7}) gives ${\psi}(\xi,\theta)$
for $\alpha (\theta)<0$.

\subsubsection{$\alpha (\theta)>0$}
In this case,
there is a contribution from the pole $t_{\ast}$.
Hence the integral is divided into the integral around
the pole and the one whose circuit of integration is 
${\tilde C}$.

The integral around the pole gives
\begin{equation}
\exp \left( i\xi \cos \frac{ \alpha (\theta) }{ \sqrt{\xi} } \right). \label{ap.8}
\end{equation}

The integral around $i\pi/2$ along the trajectory ${\tilde C}$ 
is the same as for $\alpha (\theta)<0$ 
and is given by Eq.~(\ref{ap.2}).
The only difference is the signature of $\tilde\alpha (\theta)$. 
By changing the integration variable from $u$ to $-u$, 
$\tilde\alpha(\theta)$ is replaced with  
$-\tilde\alpha(\theta)$ and the overall signature 
flips. 
As a result we find that the integration along $\tilde C$ gives 
\begin{eqnarray}
&&-\frac{1}{\sqrt{\pi}} \exp \left( i\xi+\frac{i \alpha (\theta)}{
			    (1-\Delta) \sqrt{\xi} }-\frac{i}{2} {\tilde
			    \alpha}^2 (\theta) \right) \cr
&&
\qquad \times 
{\rm Erfc} \left( -\frac{{\tilde \alpha}(\theta)}{ \sqrt{2} } 
    e^{3i\pi/4}
	   \right). \label{ap.9}
\end{eqnarray}
Integral Eq.~(\ref{ap.1}) around $-i\pi/2$ is also given by Eq.~(\ref{ap.7}).

Combining the results of subsections $a.$ and $b.$, 
adding the similar terms $\psi (\xi, -\theta)$, 
and also using the asymptotic form of $J_0 (\xi)$,
we have Eq.~(\ref{3.3b}).

\section{source at a finite distance}
Here we consider a point source at a finite distance. 
For a point source,
\begin{equation}
S=\frac{1}{(1-4G\mu)r_o} \delta (r-r_o) \delta (\theta-\pi) 
  \delta (z) e^{-i\omega t},  \label{1.2p}
\end{equation}
where $(1-4G\mu)r_o=\sqrt{-g}$. 
We consider a solution written 
in the form of the following expansion, 
\begin{equation}
\phi(r,\theta,z)=
  \sum_{m=0}^{\infty} \int_{-\infty}^{\infty}
   dk\, f_{m,k} (r)\, \cos m \theta \,e^{ikz}. 
\label{1.4p}
\end{equation}
The solution for $f_{k,m}(r)$ is the same as 
$f_m(r)$ in (\ref{1.6}) but $\omega$ contained 
in $\xi$ and $\xi_o$ 
are here replaced with $\sqrt{\omega^2-k^2}$, and 
\begin{eqnarray}
N_m =
\frac{1}{1-\Delta} \frac{ {\epsilon}_m {(-1)}^m}{8i\pi},
\label{1.7p}
\end{eqnarray}

First we compute  
$\displaystyle\phi_k(r,\theta):= \sum_m^{\infty}
  f_{m,k}(r)\cos m\theta$ for $r<r_o$. 
As in the case of Bessel function, 
we also use the integral representation for Hankel function
\begin{equation}
 H^{(1)}_\nu(z)={1\over i\pi}\int_{C_H}
  ds e^{z\sinh s -\nu s}.
\end{equation}
Here the integration is to be performed along the path 
$C_{H}$ presented in Fig.~\ref{contour}. 
Using the above formula and (\ref{3.1}), we have 
\begin{eqnarray}
\phi_k(r,\theta) 
& = & 
   \sum_m^{\infty}
  {i \epsilon_m (-1)^m \over 32\pi^3 (1-\Delta)}
 \int_{\infty-i\pi}^{\infty+i\pi} \!\!\! dt\, e^{\xi\sinh t-\nu_m t}\cr
 &&\quad\times 
 \int_{-\infty}^{\infty+i\pi}\!\!\! ds\, e^{\xi_o\sinh s-\nu_m s} 
   \left(e^{im\theta}+e^{-im\theta}\right)\cr
& \approx & 
  {i \over 16\pi^3 (1-\Delta)}
 \int_{\infty-i\pi}^{\infty+i\pi} \!\!\! dt\, e^{\xi\sinh t}\cr
&&\times
 \int_{-\infty}^{\infty+i\pi}\!\!\!\! ds\, e^{\xi_o\sinh s} 
   \Biggl({1\over 1+e^{-\frac{t+s}{1-\Delta}+i\theta-\epsilon}}\cr
&&\hspace*{3.5cm}  +(\theta\to -\theta)\Biggr).  
\end{eqnarray}
We introduce a new variable $t'\equiv t+s-i\pi/2$. 
Under the assumption that $\xi_o\gg 1$, 
the integration over $s$ is dominated by the contribution around 
$s=i\pi/2$. Hence, the integration contour for $t$ is unaltered 
even if we change the integration variable from $t$ to $t'$. 
After this change of the variable, we have
\begin{eqnarray}
\phi_k(r,\theta) 
& \approx & 
  {i \over 16\pi^3 (1-\Delta)}
 \int_{\infty-i\pi}^{\infty+i\pi} \!\!\! dt' 
 \int_{-\infty}^{\infty+i\pi} \!\!\!ds\,
e^{f(t',s)}\cr
&&\quad \times 
   \left[{1\over 1+e^{-\frac{1}{1-\Delta} (t'+i\pi/2)+i\theta-\epsilon}}
  +(\theta\to -\theta)\right], \cr
&&
\end{eqnarray}
where 
\begin{equation}
f(t',s):= \xi\sinh (t'-s+i\pi/2)
   +\xi_o\sinh s. 
\end{equation}
We expand the exponent around a zero of its derivative. 
The derivative vanishes at $s=s_0$, and $s_0$ is given by 
\begin{equation}
 \tanh s_0= {\xi\cosh(t'+i\pi/2)-\xi_o\over \xi\sinh(t'+i\pi/2)}. 
\end{equation}
Taylor expansion of $f(t',s)$ around $s=s_0$ becomes 
\begin{eqnarray}
 f(t',s)& =&i\sqrt{\xi^2+\xi_o^2-2\xi\xi_o\cosh(t'+i\pi/2)}\cr
&&\times 
  \left(1+{1\over 2}\left( s-s_0 \right)^2+\cdots\right).
\end{eqnarray}
We truncate this expansion at the quadratic order because the 
higher order terms are suppressed by $1/\xi$ or $1/\xi_o$. Performing 
gaussian integral, we obtain 
\begin{eqnarray}
\phi_k(r,\theta) 
& \approx & 
 {\sqrt{2\pi} i \over 16\pi^3 (1-\Delta)}
 \int_{\infty-i\pi}^{\infty+i\pi} dt' 
 {e^{f(t',s_0)}\over \sqrt{-f(t',s_0)}} 
\cr && \times      
\left({1\over 1+e^{-\frac{1}{1-\Delta} (t'+i\pi/2)+i\theta-\epsilon}}
  +(\theta\to -\theta)\right).\cr
&&  
\end{eqnarray}
Further, we expand $f(t',s_0)$ around an approximate 
stationary point at $t'=i\pi/2$. Then we have 
\begin{equation}
 f(t',s_0)=i(\xi+\xi_o)+{i\xi\xi_o\over 2(\xi+\xi_o)}
    \left(t'-i{\pi\over 2}\right)^2+\cdots. 
\end{equation}
Again we truncate this expansion at the 
quadratic order for the same reason as before. 
Then one finds that $\phi_k(r,\theta)$ is approximately 
given by 
\begin{eqnarray}
\phi_k(r,\theta) 
\approx 
  -{\sqrt{2\pi} e^{i\pi/4}
   \over 8\pi^2 
    \sqrt{\xi+\xi_o}}
   \tilde {\phi}_k(r,\theta),
\end{eqnarray}
where
\begin{eqnarray}
&&{\tilde {\phi}_k} (r,\theta)=e^{i (\xi+\xi_o)} 
{\cal F} \left(  \frac{\xi
  \xi_o}{\xi+\xi_o}, \theta \right),
\end{eqnarray}
and
\begin{eqnarray}
{\cal F}(x,\theta) &:=& {1\over 2i\pi(1-\Delta)}
    \int_{\infty-i\pi}^{\infty+i\pi} \!\!\!\! dt'\,
e^{ \frac{i x}{2} (t'-i{\pi\over 2})^2} \cr
&&\times
   \left({1\over 1-e^{-\frac{1}{1-\Delta} (t'+i\pi/2)+i\theta-\epsilon}}
  +(\theta\to -\theta)\right). \cr&&\label{b1}
\end{eqnarray}
The function $\tilde\phi(r,\theta)$
is almost identical to $\phi(r,\theta)$ 
discussed in Sec.III, 
except that $e^{i\xi}$ and other $\xi$ are replaced with 
$e^{i(\xi+\xi_o)}$ and $\frac{\xi \xi_o}{\xi+\xi_o}$, 
respectively. 

Finally, we perform the integration over $k$. 
From (\ref{b1}),
we have 
\begin{equation}
\phi(r,\theta,z)
\approx  
 -{\sqrt{2\pi} e^{i\pi/4}
   \over 8\pi^2 }
\int dk {e^{i(\xi+\xi_o)}\over\sqrt{\xi+\xi_o}} e^{ikz} 
   {\cal F} \left(  \frac{\xi
  \xi_o}{\xi+\xi_o}, \theta \right).
\end{equation}
Since $\xi+\xi_o=\sqrt{\omega^2-k^2}(r+r_o)$, 
we can invoke the saddle point method again to 
perform $k$-integral when $r+r_o$ is large. 
Evaluating the contribution from the saddle point at 
$k=\omega z/D$ with $D\equiv \sqrt{z^2+(r+r_o)^2}$, we obtain 
\begin{equation} 
\phi(r,\theta,z)
   \approx -{1
   \over 4\pi D} 
     e^{i\omega D}
   {\cal F}\left( \frac{\omega r r_o}{D},\theta \right).
\end{equation}
The calculation for $r>r_o$ can be done 
in a completely parallel way, and the final result
becomes identical to the case with $r<r_o$.


\begin{thebibliography}{99}
\bibitem{Ohanian}
H.~C.~Ohanian, Phys.\ Rev.\ D {\bf 8}, 2734 (1973),
H.~C.~Ohanian, Int.~J.~Theor.~Phys. {\bf 9}, 425 (1974),
H.~C.~Ohanian, Astrophys.~J. {\bf 271}, 551 (1983).

\bibitem{Bliokh}
P.~V.~Bliokh and A.~A.~Minakov, Ap\& SS. {\bf 34}, L7 (1975).

\bibitem{Bontz}
R.~J.~Bontz and M.~P.~Haugan, Ap\& SS. {\bf 78}, 199 (1981)

\bibitem{Thorne}
K.~S.~Thorne, in Gravitational Radiation, ed. N.~Deruell and T.~Piran (Amsterdam: North-Holland), 28, (1983)

\bibitem{Deguchi}
S.~Deguchi and W.~D.~Watson, Astrophys.~J. {\bf 307}, 30 (1986).

\bibitem{Takahashi:2003ix}
R.~Takahashi and T.~Nakamura,
Astrophys.\ J.\  {\bf 595}, 1039 (2003)
[arXiv:astro-ph/0305055].

\bibitem{Seto:2003iw}
N.~Seto,
Phys.\ Rev.\ D {\bf 69}, 022002 (2004)
[arXiv:astro-ph/0305605].

\bibitem{Nakamura:1997sw}
T.~T.~Nakamura,
Phys.\ Rev.\ Lett.\  {\bf 80}, 1138 (1998).

\bibitem{Yamamoto:2003cd}
K.~Yamamoto,
arXiv:astro-ph/0309696.

\bibitem{Baraldo:1999ny}
C.~Baraldo, A.~Hosoya and T.~T.~Nakamura,
D {\bf 59}, 083001 (1999).

\bibitem{T.Nakamura:1999}
T.~T.~Nakamura and S.~Deguchi
Prog. Theor. Phys. Suppl. {\bf 133}, 137 (1999).

\bibitem{Yamamoto:2003wg}
K.~Yamamoto and K.~Tsunoda,
Phys.\ Rev.\ D {\bf 68}, 041302 (2003)
[arXiv:astro-ph/0309694].

\bibitem{Suyama:2005mx}
T.~Suyama, R.~Takahashi and S.~Michikoshi,
arXiv:astro-ph/0505023.

\bibitem{Takahashi:2005sx}
R.~Takahashi, T.~Suyama and S.~Michikoshi,
Astron.\ Astrophys.\ {\bf 438}, L5 (2005)
[arXiv:astro-ph/0503343].

\bibitem{Takahashi:2004mc}
R.~Takahashi,
Astron.\ Astrophys.\ {\bf 423}, 787 (2004)
[arXiv:astro-ph/0402165].

\bibitem{Hindmarsh:1994re}
M.~B.~Hindmarsh and T.~W.~B.~Kibble,
Rept.\ Prog.\ Phys.\  {\bf 58}, 477 (1995)
[arXiv:hep-ph/9411342].

\bibitem{vilenkin}
A.~Vilenkin and E.~P.~S.~Shellard,
Cosmic Strings and Other Topological Defects,
Cambridge University Press (Cambridge, 2000).

\bibitem{Dvali:1998pa}
G.~R.~Dvali and S.~H.~H.~Tye,
arXiv:hep-ph/9812483.

\bibitem{Dvali:1999tq}
G.~R.~Dvali,
B {\bf 459}, 489 (1999)
[arXiv:hep-ph/9905204].

\bibitem{Burgess:2001fx}
C.~P.~Burgess, M.~Majumdar, D.~Nolte, F.~Quevedo, G.~Rajesh and R.~J.~Zhang,
 {\bf 0107}, 047 (2001)
[arXiv:hep-th/0105204].

\bibitem{Alexander:2001ks}
S.~H.~S.~Alexander,
Phys.\ Rev.\ D {\bf 65}, 023507 (2002)
[arXiv:hep-th/0105032].

\bibitem{Dvali:2001fw}
G.~R.~Dvali, Q.~Shafi and S.~Solganik,
arXiv:hep-th/0105203.

\bibitem{Jones:2002cv}
N.~Jones, H.~Stoica and S.~H.~H.~Tye,
 {\bf 0207}, 051 (2002)
[arXiv:hep-th/0203163].

\bibitem{Shiu:2001sy}
G.~Shiu and S.~H.~H.~Tye,
Phys.\ Lett.\ B {\bf 516}, 421 (2001)
[arXiv:hep-th/0106274].

\bibitem{Majumdar:2002hy}
M.~Majumdar and A.~Christine-Davis,
JHEP {\bf 0203}, 056 (2002)
[arXiv:hep-th/0202148].

\bibitem{Dvali:2002fi}
G.~Dvali and A.~Vilenkin,
Phys.\ Rev.\ D {\bf 67}, 046002 (2003)
[arXiv:hep-th/0209217].

\bibitem{Jones:2003da}
N.~T.~Jones, H.~Stoica and S.~H.~H.~Tye,
Phys.\ Lett.\ B {\bf 563}, 6 (2003)
[arXiv:hep-th/0303269].

\bibitem{Dvali:2003zj}
G.~Dvali and A.~Vilenkin,
JCAP {\bf 0403}, 010 (2004)
[arXiv:hep-th/0312007].

\bibitem{Copeland:2003bj}
E.~J.~Copeland, R.~C.~Myers and J.~Polchinski,
JHEP {\bf 0406}, 013 (2004)
[arXiv:hep-th/0312067].

\bibitem{Spergel:2003cb}
D.~N.~Spergel {\it et al.}  [WMAP Collaboration],
Astrophys.\ J.\ Suppl.\  {\bf 148}, 175 (2003)
[arXiv:astro-ph/0302209].

\bibitem{Percival:2002gq}
W.~J.~Percival {\it et al.}  [The 2dFGRS Team Collaboration],
Mon.\ Not.\ Roy.\ Astron.\ Soc.\  {\bf 337}, 1068 (2002)
[arXiv:astro-ph/0206256].

\bibitem{Jeong:2004ut}
E.~Jeong and G.~F.~Smoot,
Astrophys.\ J.\  {\bf 624}, 21 (2005)
[arXiv:astro-ph/0406432].

\bibitem{Pogosian:2003mz}
L.~Pogosian, S.~H.~H.~Tye, I.~Wasserman and M.~Wyman,
Phys.\ Rev.\ D {\bf 68}, 023506 (2003)
[arXiv:hep-th/0304188].

\bibitem{Pogosian:2004ny}
L.~Pogosian, M.~C.~Wyman and I.~Wasserman,
arXiv:astro-ph/0403268.

\bibitem{Wyman:2005tu}
M.~Wyman, L.~Pogosian and I.~Wasserman,
Phys.\ Rev.\ D {\bf 72}, 023513 (2005)
[arXiv:astro-ph/0503364].

\bibitem{Kaspi:1994hp}
V.~M.~Kaspi, J.~H.~Taylor and M.~F.~Ryba,
Astrophys.\ J.\  {\bf 428} (1994) 713.

\bibitem{Thorsett:1996dr}
S.~E.~Thorsett and R.~J.~Dewey,
Phys.\ Rev.\ D {\bf 53}, 3468 (1996).

\bibitem{McHugh:1996hd}
M.~P.~McHugh, G.~Zalamansky, F.~Vernotte and E.~Lantz,
Phys.\ Rev.\ D {\bf 54}, 5993 (1996).

\bibitem{Lommen:2002je}
A.~N.~Lommen,
arXiv:astro-ph/0208572.

\bibitem{Bouchet:2000hd}
F.~R.~Bouchet, P.~Peter, A.~Riazuelo and M.~Sakellariadou,
Phys.\ Rev.\ D {\bf 65}, 021301 (2002)
[arXiv:astro-ph/0005022].

\bibitem{Rocher:2004my}
J.~Rocher and M.~Sakellariadou,
Phys.\ Rev.\ Lett.\  {\bf 94}, 011303 (2005)
[arXiv:hep-ph/0412143].

\bibitem{Sazhin:2005fd}
M.~Sazhin, M.~Capaccioli, G.~Longo, M.~Paolillo and O.~Kovanskaya,
arXiv:astro-ph/0506400.

\bibitem{Sazhin:2003cp}
M.~Sazhin {\it et al.},
Mon.\ Not.\ Roy.\ Astron.\ Soc.\  {\bf 343}, 353 (2003)
[arXiv:astro-ph/0302547].

\bibitem{Linet:1986db}
B.~Linet,
Annales Poincare Phys.\ Theor.\  {\bf 45}, 249 (1986).

\bibitem{Vilenkin:1981zs}
A.~Vilenkin,
Phys.\ Rev.\ D {\bf 23}, 852 (1981).

\bibitem{Gott:1984ef}
J.~R.~I.~Gott,
Astrophys.\ J.\  {\bf 288}, 422 (1985).

\bibitem{Schneider}
P.~Schneider, J.~Ehlers and E.~E.~Falco,
Gravitational Lenses,
Springer,
New York, (1992)

\bibitem{lisa}
See http://lisa.jpl.nasa.gov/

\bibitem{skn01}
N.~Seto, S.~Kawamura and T.~Nakamura,
Phys.\ Rev.\ Lett.\  {\bf 87}, 221103 (2001).

\bibitem{bbo}
See http://universe.nasa.gov/program/bbo.html

\bibitem{Cutler:1994ys}
C.~Cutler and E.~E.~Flanagan,
Phys.\ Rev.\ D {\bf 49}, 2658 (1994)
[arXiv:gr-qc/9402014].

\bibitem{Albrecht:1989mk}
A.~Albrecht and N.~Turok,
Phys.\ Rev.\ D {\bf 40}, 973 (1989).

\bibitem{Bennett:1989yp}
D.~P.~Bennett and F.~R.~Bouchet,
Phys.\ Rev.\ D {\bf 41}, 2408 (1990).

\bibitem{Allen:1990tv}
B.~Allen and E.~P.~S.~Shellard,
Phys.\ Rev.\ Lett.\  {\bf 64}, 119 (1990).

\bibitem{Matzner}
R.~A.~Matzner,
Comput.\ Phys.\ {\bf 2}, 51 (1989).

\bibitem{Jackson:2004zg}
M.~G.~Jackson, N.~T.~Jones and J.~Polchinski,
arXiv:hep-th/0405229.

\bibitem{Haehnelt:1994wt}
M.~G.~Haehnelt,
Mon.\ Not.\ Roy.\ Astron.\ Soc.\  {\bf 269}, 199 (1994)
[arXiv:astro-ph/9405032].

\bibitem{Islam:2003nt}
R.~R.~Islam, J.~E.~Taylor and J.~Silk,
Mon.\ Not.\ Roy.\ Astron.\ Soc.\  {\bf 354}, 629 (2004)
[arXiv:astro-ph/0309559].

\bibitem{Ioka:2005pm}
K.~Ioka and P.~Meszaros,
arXiv:astro-ph/0502437.

\bibitem{Ruffa:1999}
A.~A.~Ruffa,
Astrophys.\ J.\  {\bf 517}, L31 (1999)

\bibitem{Vilenkin:2005jg}
A.~Vilenkin,
arXiv:hep-th/0508135.

\bibitem{Siemens:2001dx}
X.~Siemens and K.~D.~Olum,
Nucl.\ Phys.\ B {\bf 611}, 125 (2001)
[Erratum-ibid.\ B {\bf 645}, 367 (2002)]
[arXiv:gr-qc/0104085].

\bibitem{Siemens:2002dj}
X.~Siemens, K.~D.~Olum and A.~Vilenkin,
Phys.\ Rev.\ D {\bf 66}, 043501 (2002)
[arXiv:gr-qc/0203006].
\end{thebibliography}
\end{document}